\documentclass{aa}
\usepackage{graphicx}
\usepackage{subfig}

\usepackage{txfonts}
\newcommand{\oii}{[O\,{\sc{ii}}]}
\def\ha{H$\alpha$}
\def\hb{H$\beta$}

\begin{document}

   \title{The OTELO survey}
      \subtitle{New evidence of downsizing from the specific star formation rates, stellar mass functions, and star formation histories of a sample of low-mass galaxies at $0.38<z<1.43$}
   \author{Bernab\'e Cedr\'es
          \inst{1,2,3}
   \and{\'Angel Bongiovanni}
          \inst{4,3}
   \and{Jordi Cepa}
          \inst{1,2,3}
   \and{Carmen P. Padilla-Torres}
          \inst{1,2,5,3}
   \and{Jakub Nadolny}
          \inst{6,1}
   \and{Maritza A. Lara-L\'opez}
          \inst{7}
   \and{J. Ignacio Gonz\'alez-Serrano}
          \inst{8,3}
   \and{Emilio J. Alfaro}
          \inst{9}
   \and{Miguel Cervi\~no}
          \inst{10}
   \and{Jes\'us Gallego}
          \inst{7}
   \and{Mauro Gonz\'alez-Otero}
          \inst{3}
   \and{Miguel S\'anchez-Portal}
          \inst{4,3}
   \and{Jos\'e A. de Diego}
          \inst{11}
   \and{Ricardo P\'erez-Mart\'inez}
          \inst{12,3}
   \and{Ana Mar\'ia P\'erez-Garc\'ia}
          \inst{12,3}
   \and{Mirjana Povi\'c}
          \inst{13,9,14}
          }
   \institute{Instituto de Astrof{\'i}sica de Canarias (IAC), 38200 La Laguna, Tenerife, Spain\\
              \email{bcedres@iac.es}
         \and
             Departamento de Astrof{\'i}sica, Universidad de La Laguna (ULL), 38205 La Laguna, Tenerife, Spain
        \and
            Asociaci\'on Astrof{\'i}sica para la Promoci\'on de la Investigaci\'on, Instrumentaci\'on y su Desarrollo, ASPID, 38205 La Laguna, Tenerife, Spain
        \and
        Institut de Radioastronomie Millim\'etrique (IRAM), Av. Divina Pastora 7, N\'ucleo Central 18012, Granada, Spain
        \and
        Fundaci\'on Galileo Galilei--INAF, Rambla Jos\'e Ana Fern\'andez P\'erez, 7, 38712 Bre\~na Baja, Tenerife, Spain
        \and
        Astronomical Observatory Institute, Faculty of Physics and Astronomy, Adam Mickiewicz University, ul.~S{\l}oneczna 36, 60-286 Pozna{\'n}, Poland 
        \and
        Departamento de F\'isica de la Tierra y Astrof\'isica, Instituto de F\'isica de Part\'iculas y del Cosmos, IPARCOS. Universidad Complutense de Madrid (UCM), 28040 Madrid, Spain
        \and
        Instituto de F\'isica de Cantabria (CSIC-Universidad de Cantabria), 39005 Santander, Spain
        \and
        Instituto de Astrof\'isica de Andaluc\'ia (CSIC), 18080 Granada, Spain
        \and
        Centro de Astrobiolog\'ia (CSIC/INTA), ESAC Campus, 28692 Villanueva de la Ca\~nada, Madrid, Spain
        \and
        Instituto de Astronom\'ia, Universidad Nacional Aut\'onoma de M\'exico, Apdo. Postal 70-264, 04510 Ciudad de M\'exico, M\'exico
        \and
        ISDEFE for European Space Astronomy Centre (ESAC)/ESA, PO Box 78, 28690 Villanueva de la Ca\~nada, Madrid, Spain
        \and
        Space Science and Geospatial Institute (SSGI), Entoto Observatory and Research Center (EORC), Astronomy and Astrophysics Research Division, PO Box 33679, Addis Abbaba, Ethiopia
        \and
        Physics Department, Mbarara University of Science and Technology (MUST), Mbarara, Uganda
    }
   \date{}

 
  \abstract
   {}
   {We present an analysis of the emitters (\ha, \hb, and \oii) from the OTELO survey, in order to characterize the star formation properties of low-mass galaxies ($<10^9$ M$_{\odot}$ stellar masses).}
   {We calculated the specific star formation rate function, the stellar mass function, and, by integrating them, the associated densities for both quantities: the specific star formation rate density and the stellar mass density. We obtained the star formation history of our low-mass sample galaxies by fitting the spectral energy distribution of the galaxies. We also compared our results with those from the literature at different mass regimes and redshifts.}
   {The specific star formation rate density and the stellar mass density for low-mass galaxies do not depend on the redshift, contrary to the behaviour presented by the high-mass galaxies. We found that the star formation histories of low-mass galaxies are characterized by a constant star formation rate, in contrast to high-mass galaxies. We interpret these results, in the context of the downsizing effect, as representative of the faster evolution of massive galaxies compared with low-mass ones.}
   {}


   \maketitle
%

\section{Introduction}
The role of the low-mass galaxies ($\log(M_{\mathrm{s}}/\mathrm{M_{\odot}})<9$, where $M_{\mathrm s}$ is the stellar mass) in the canonical evolution of star formation is of paramount importance. Following the ‘bottom-up’ paradigm, low-mass galaxies should be the dominant class of galaxies in the earlier Universe. Moreover, they could be the main drivers of reionization (see, for example, \citealt{simmonds2024} and \citealt{sales2022} and references therein). 

\noindent
\indent There is a tight correlation between the star formation rate (SFR), the metallicity, and the stellar mass of galaxies (\citealt{maritza2010}). This indicates that an important part of the evolution of galaxies may be determined by their stellar masses. 

\noindent
\indent The downsizing effect, first defined by \cite{cowie1996}, can be expressed in several ways: an earlier assembling of massive galaxies, older ages for high-mass galaxies, or as an earlier, more important star formation process for high-mass galaxies compared with low-mass ones. According to \cite{fontanot2009}, all these definitions may be equivalent. However, in this paper, we stick to the latter definition because our main goal is to study the star formation processes, not the ages or the dynamic evolutionary stage.

\noindent
\indent \cite{Hopkins2004} presented a compilation from different authors in the literature that shows the density of the star formation rate (SFRD) of galaxies, up to $z\sim6$. As was shown in that work, the SFRD increases with redshift up to a maximum at about $z\sim2$, the so-called ‘cosmic noon’. The same behaviour is reproduced in the \cite{madau2014} review, but in this case up to $z\sim8$. In that work, the authors concluded that, at its peak, the galaxies formed stars at a rate nine times higher than in present times. \cite{picouet2023} reproduced the same trend with a sample of one million galaxies in the HSC-CLAUDS survey (\citealt{desprez2023}) and found a plateau between $1\le z \le 2$ instead of the gradual decline in SFRD reported in \cite{madau2014}. However, in all these studies the results were obtained for intermediate- and high-mass galaxies ($>10^{9}\,\mathrm{M_\odot}$). On the other hand, \cite{beliletter} suggest that the SFRD in low-mass galaxies is constant with up to a redshift of $z\sim1.5$. 

\noindent
\indent Less work has been done regarding the specific star formation rate (sSFR). The sSFR is defined as $sSFR=SFR/M_{s}$, where $M_{\rm s}$ is the stellar mass of the galaxy. \cite{ilbert2015} studied the median values of the sSFR and found that it evolves with the redshift and the stellar mass: the sSFR increased with redshift and decreased with stellar mass. On the other hand, \cite{lehnert} reported a decrease in the sSFR for $z\le2$ and suggested that this should be due to the gas being depleted below the threshold to maintain the star formation. However, both studies only reach down to $\log(M_{\mathrm{s}}/\mathrm{M_{\odot}})=9.5$, and thus they did not include low-mass galaxies.

\noindent
\indent There have been numerous studies on the evolution of the stellar mass function ($\phi(M_{\mathrm{S}}$), SMF), as well as the evolution of the stellar mass density ($\rho_*$), for star-forming galaxies (SFGs), quiescent galaxies, or both types simultaneously (\citealt{perezgonzalez2008}, \citealt{wilkins2008}, \citealt{marchesini2009}, \citealt{muzzin2013}, \citealt{ilbert2013}, \citealt{sobral2014}, \citealt{kikuchihara2020}, and \citealt{weaver2023}, among many others). 

\noindent
\indent \cite{mimi2016} found that the number density of high-mass galaxies at high redshift ($4<z<8$) presents a steeper slope than galaxies with masses lower than  $\log(M_{\mathrm{s}}/\mathrm{M_{\odot}})<10$, with a completeness limit at about  $\log(M_{\mathrm{s}}/\mathrm{M_{\odot}})\sim9$ . At the same time, the authors suggest the opposite behaviour for redshifts lower than $z<4$.

\noindent
\indent On a similar note, \cite{ilbert2013}, in the range of $9.5<\log(M_{\mathrm{s}}/\mathrm{M_{\odot}})<13$, found a clear evolution of the SMF from $1 \leq z \leq 4$ of about 1\,dex for SFGs, but with a slower evolution for $z \leq 1$. They reached $\log(M_{\mathrm{s}}/\mathrm{M_{\odot}})<8.25$, but only at redshift $z\sim0.2$. They also found that galaxies with lower masses evolved more rapidly than the most massive galaxies. On the other hand, \cite{sobral2014} found no evolution at all, at least for the faint-end slope of the SMFs for emission line galaxies (ELGs) at low to intermediate redshifts ($0.40<z<2.23$), detecting stellar masses as low as $\log(M_{\mathrm{s}}/\mathrm{M_{\odot}})\sim9.0$ at $z\sim0.4$, and about $\log(M_{\mathrm{s}}/\mathrm{M_{\odot}})\sim9.75$ at $z\geq0.84$.

\noindent
\indent The OTELO (OSIRIS
Tunable Filter Emission Line Object) survey offers the perfect opportunity to fill the gap. We present a study of a sample of low-mass galaxies, in order to shed light on the behaviour of low-mass galaxies at intermediate redshifts.

\noindent
\indent In Sect. \ref{sec2}, we describe the OTELO survey and present the stellar mass distribution of the emitters at redshifts 0.38, 0.88, and 1.43. In Sect. \ref{sec3}, we present the sSFRFs of the emitters, calculate the specific SFRDs ($\rho_{\mathrm{sSFR}}$), and compare our results with those from the literature. In Sect. \ref{sec4}, we repeat the same process as in Sect. \ref{sec3}, but in this case with the stellar mass of galaxies, calculating the SMF and the $\rho_*$. In Sect. \ref{sec5}, we explore the star formation histories (SFHs) of the galaxies by fitting their spectral energy distribution (SED) through the CIGALE code, and analyse the derived quantities as a function of the stellar mass. Finally, in Sect. \ref{sec6} we present the main conclusions of this work.

\noindent
\indent Throughout this paper, we assume a standard $\Lambda$CDM cosmology with $\Omega_\Lambda = 0.7$, $\Omega_{\mathrm{m}} = 0.3$, and $\mathrm{H}_0 = 70 \,\mathrm{km}\,  \mathrm{s}^{-1}\, \mathrm{Mpc}^{-1}$. We also assume a \cite{kroupaimf} initial mass function (IMF), unless stated otherwise.

\section{The OTELO survey}\label{sec2}
As is described in detail in \cite{otelo}, the OTELO survey is a blind, pencil-beam survey of the Extended Groth Strip field, based on 2D spectroscopic techniques, employing the tunable filters of the OSIRIS instrument at Gran Telescopio Canarias (GTC). It covers 56\,arcmin$^2$ and has a spectral resolution of $R\sim700$. It samples the wavelength region between 8950--9300\,\AA\ in an atmospheric window. 
\noindent
The total number of raw sources candidates detected in the survey is 11237 by aperture photometry using {\tt SEXtractor} (\citealt{bertin1996}).

Employing the ancillary photometric data, also described in \cite{otelo}, the SED of each source was fitted through the LePhare code (\citealt{arnouts1999}; \citealt{ilbert2006}), providing photometric redshifts for 6600 sources, among other solutions.

\subsection{The pseudo-spectra and the inverse convolution}
\begin{figure}
    \centering
    \includegraphics[width=\hsize]{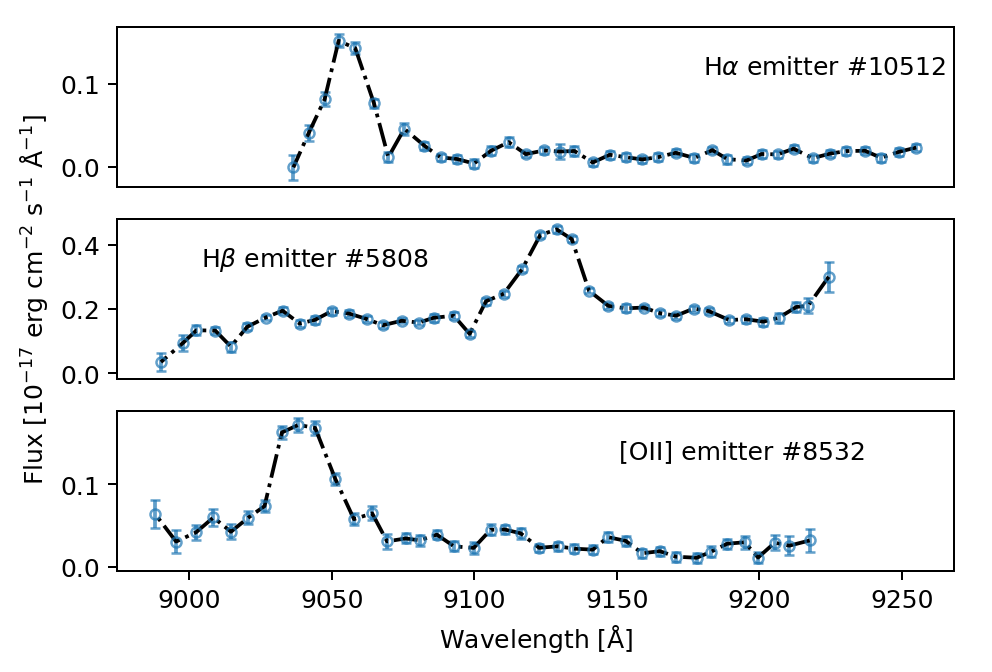}
    \caption{Pseudo-spectra for selected \ha, \hb, and \oii\ emitters from the OTELO survey.}
    \label{pseudo}
\end{figure}

Due to the design of the tunable filters installed at OSIRIS instrument, for each of the sources, one of the products obtained is not a spectrum but the result of the convolution in wavelength space of the SED of the source and the special instrumental response, or pseudo-spectra. In this case, such an instrumental response is a succession of Airy profiles. As an example of selected pseudo-spectra, a subsample of emitters employed in this work is represented in Fig. \ref{pseudo}

To extract information from the pseudo--spectra, we developed the ‘inverse convolution’ method, described in detail by \cite{jakub2020}. This method is based on $10^6$ Monte Carlo simulations, sampling the probability distribution function (PDF) of the intensity of the fluxes, the shape of the main emission and absorption features, the continuum, and the redshift.

\subsection{Redshift determination, selection of the samples, and extinction correction}
In the articles presenting the subsamples for \ha, \hb, and \oii$\lambda3727$ (hereafter \oii) emitters (\citealt{marina2019}, \citealt{rocio2021}, and \citealt{beli2021}, respectively), the process of selection is described in detail. As a summary, at first, sources with a $z_{\rm best}$ in the range of the expected redshift (0.40 for \ha, 0.90 for \hb, and 1.43 for \oii) were selected. The variable $z_{\rm best}$ is defined in \cite{bongio2020} as the z$\_$BEST$\_$deepY (the photo--$z$ solution from LePhare code). With this, 202 candidates for the \ha\ emitters, 87 for the \hb\ emitters, and 332 for the \oii\ emitters were obtained. Then, at least three researchers selected a new redshift, $z_{\rm guess}$, by visual inspection. This new value was assigned to the peak of the emission line observed in the pseudo-spectrum. This value was given to the inverse deconvolution software to be used as a prior to determine the final redshift.

The reddening, $E(B-V)$, for all the emitters, was obtained as an output from the fitting using the LePhare code. Employing that reddening, the extinction through the law in \cite{calzetti2000} was obtained. The inherent extinction of the fitting templates employed (\citealt{kinney1996}) was also taken into account.

\subsection{Completeness of the samples}
\begin{figure}
    \centering
    \includegraphics[width=\hsize]{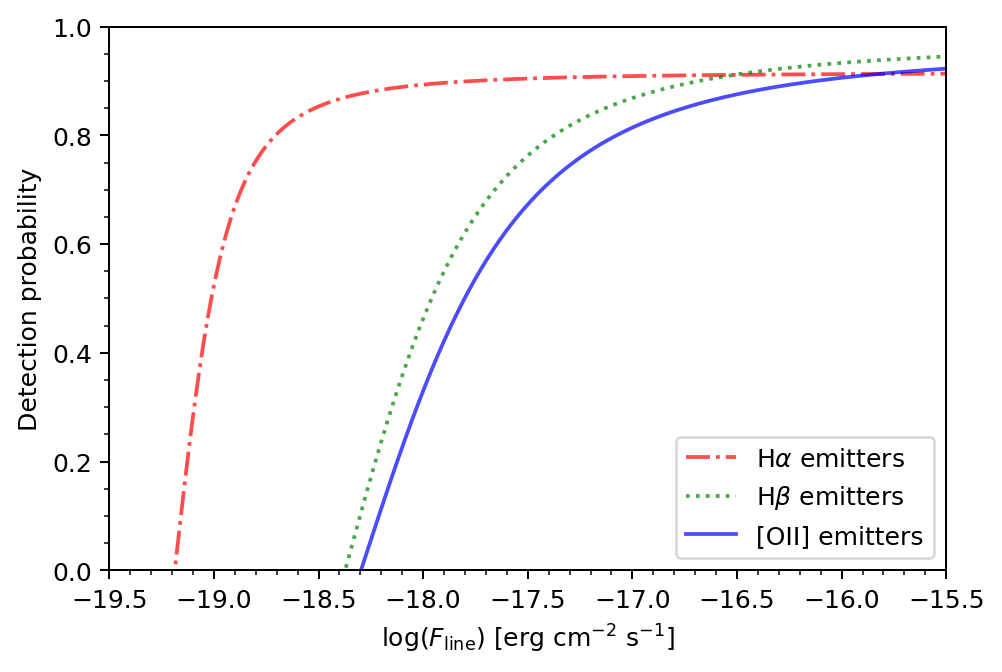}
    \caption{Mean detection probability for the emitters at OTELO.}
    \label{complet}
\end{figure}

The completeness of the OTELO sample is discussed in \cite{bongio2020}. The full sample has a 50\% completeness at $\mathrm{AB}=26.5$. The subsamples for \ha, \hb, and \oii\ are described in \cite{marina2019}, \cite{rocio2021}, and \cite{beli2021}, respectively. Those values were obtained by simulations of synthetic pseudo-spectra. In Fig. \ref{complet}, the detection probability of an \ha, \hb, and \oii\ detection line (dash-dotted red line, dotted green line, and solid blue line, respectively) is represented based on their fluxes.

In total, the final subsamples obtained are: 46 \ha\ emitters at $z\sim0.40$ (\citealt{marina2019}), 40 \hb\ emitters at $z\sim0.90$ (\citealt{rocio2021}), and 60 \oii\ emitters at $z\sim1.43$ (\citealt{beli2021}). In Table \ref{resgal} are summarized the characteristics of these emitters, including the redshift, the number of emitters, lower and upper limits on mass, and the mean luminosity of the sources.

\begin{table}[]
    \centering
    \caption{Main properties of the ELGs from the OTELO survey.}
    \begin{tabular}{c|c|c|c|c|c}
    \hline
        Source & $\langle z\rangle$ & Number of & $\log M_{\mathrm{min}}$ & $\log M_{\mathrm{max}}$ & $\langle L \rangle$ \\
         & & emitters & [M$_\odot$] & [M$_\odot$] & [erg s$^{-1}$]\\
    \hline
    \ha\ & 0.38 & 46 & 7.12 & 9.46 & 3.88$\times 10^{40}$ \\
    \hb\ & 0.88 & 40 & 7.82 & 10.70 & 2.83$\times 10^{40}$\\
    \oii\ & 1.43 & 60 & 7.90 & 10.93 & 2.45$\times 10^{41}$\\
    \hline
    \end{tabular}
    \label{resgal}
\end{table}

\subsection{The stellar mass distribution of the emitters}
The stellar masses were derived in \cite{jakub2020} following \cite{lopezsanjuan2019} recipe. As is stated in \cite{beliletter}, this kind of pencil-beam survey favours the detection of low-mass galaxies. 

\begin{figure}
    \centering
    \includegraphics[width=\hsize]{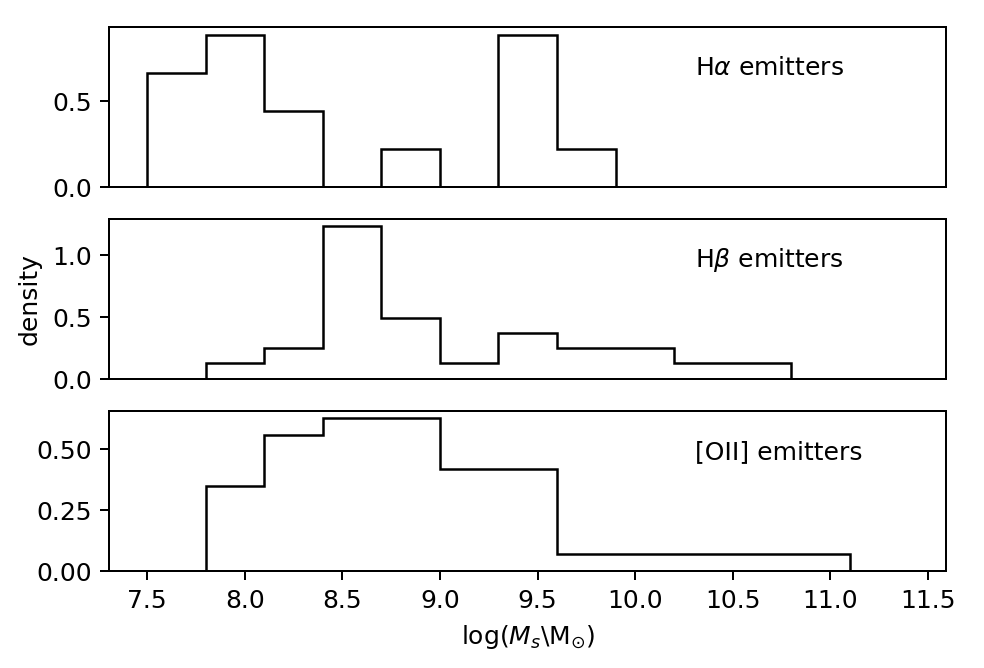}
    \caption{Distribution of the stellar mass for the ELGs employed in this study.}
    \label{masaem}
\end{figure}

If we focus on the ELGs detected in OTELO, in the range $0.38\leq z \leq1.42$, the results are the ones shown in Fig. \ref{masaem}. From a total of 146 emitters considered here, 82\%  have stellar masses of $\log(M_*/\mathrm{M_{\odot}})\leq9.5$ and 66\% are below $\log(M_*/\mathrm{M_{\odot}})\leq9$. Thus, the majority of the ELGs from our sample are low-mass galaxies.

\section{The specific star formation rate function}\label{sec3}

\subsection{The star formation rate of the OTELO survey}
In \cite{beli2021} and \cite{beliletter} the star formation rate functions (SFRFs) and SFRDs for the \ha, \hb, and \oii\, emitters in OTELO are reported. The SFR was calculated in the following way for the \ha\ and \hb\ emitters, employing the equations proposed by \cite{kennicutt1998}:
\begin{equation}
 SFR(\mathrm(M_{\odot}yr^{-1})=5.29\times10^{-42}\,L(H\alpha)\,(\mathrm{erg\,s^{-1}})   
\end{equation}
\begin{equation}
 SFR(\mathrm{M_{\odot} yr^{-1}})=1.057\times10^{-41}\,L(H\beta)\,(\mathrm{erg\,s^{-1}}),   
\end{equation}
where $L$ is the luminosity (corrected for extinction) for each line.

For the \oii\ lines, the SFR was calculated employing the calibration proposed by \cite{kewley2004}:
\begin{equation}
    SFR(\mathrm{M_\odot\,yr^{-1}})=\frac{5.29\times10^{-42}\,L(\mathrm{[OII]})\,(\mathrm{erg\, s^{-1}})}{-1.75\times[\log(\mathrm{O/H})+12]+16.73},
\end{equation}
where $L$ is the luminosity in \oii\ corrected for extinction, and $\log(\mathrm{O/H})+12$ is the metallicity for each emitter. This was done this way in order to take into account the dependence on the abundance of the SFR from \oii\ lines. There, a mean metallicity value of $\sim8.5$ was assumed for each galaxy (\citealt{henry2013}). The results presented in \cite{beliletter} are compatible with a lack of evolution in the SFRD with the redshift for low-mass galaxies.

\subsection{Building the specific star formation rate function}{\label{sec:ssfr}}
The sSFR was first introduced in \cite{guzman1997}. It is defined as the SFR per unit of stellar mass. As \cite{katsianis2021} indicate, it measures both the rate of star formation and its efficiency. 
The sSFRF ($\phi(sSFR)$) is defined as the number of galaxies per unit of sSFR and per unit of volume. \cite{ilbert2015} proposes the study of the sSFRF as a way to take into account the completeness effects as well as to obtain information on the distribution of the galaxies around the median of the sSFR. 

We can define the value of $\phi(sSFR)$ in the same way as for the luminosity function in \cite{bongio2020}:
\begin{center}
    \begin{equation}
        \label{phi}
        \phi[\log (sSFR)]= \frac{4\pi}{\Omega}\Delta[\log (sSFR)]^{-1}\sum_\ensuremath{i} \frac{1}{V_\ensuremath{i} d_\ensuremath{i}},
    \end{equation}
\end{center}
\noindent where $\Omega$ is the surveyed solid angle of OTELO survey ($\sim4.7\times10^{-6}$\,str), $V_i$ is the co-moving volume for the $i$ source, $\Delta[\log(sSFR)]=0.15$ is the adopted binning, and $d_{i}$ is the detection probability for the $i$th galaxy, and this probability is an indication of the completeness of the sample. The values of $d_{i}$ are the ones presented in Fig. \ref{complet}. To compare our results with the results from the bibliography, we only have included emitters with masses up to $\log(M_*/\mathrm{M_{\odot}})<9$.

The volumes covered by OTELO at redshifts 0.38, 0.88, and 1.43 are $1.40\times10^3$ Mpc$^3$, $5.19\times10^3$ Mpc$^3$, and $10.21\times10^3$ Mpc$^3$, respectively. These small co-moving volumes introduce significant sensitivity to cosmic variance (CV), a factor that has been shown in larger surveys such as COSMOS and GOODS to bias density estimates. This effect necessitates a careful correction that is summarized in \cite{marina2019}, \cite{rocio2021}, and \cite{beli2021} for the \ha, \hb, and \oii\, emitters, respectively. There, the recipe presented in \cite{somerville2004} to calculate the CV was employed.

\noindent Following the prescription of \cite{ilbert2015}, we can fit the sSFRF with a log-normal function with the form
\begin{equation}
    \phi(sSFR)=\frac{\Phi^*}{\sigma\sqrt{2\pi}}\exp{\left (-\frac{\log^2(sSFR/sSFR^*)}{2\sigma^2}\right),}
    \label{lognorm}
\end{equation}
where $\Phi^*$ is a normalization factor, $sSFR^*$ is the characteristic sSFR, and $\sigma$ is a dimensionless standard deviation.

The fitting of eq. \ref{lognorm} was obtained by employing a least-squares minimization algorithm based on the Levenberg--Marquardt method. After getting the guess values of the parameters, a Monte Carlo algorithm was carried out, repeating the fit $10^5$ times, a number large enough to obtain stable results, and small enough to have the results in reasonable computing times. For each repetition, the values of $\phi[\log (sSFR)]$ were randomly freed to have any value inside the uncertainties. Then, the standard deviation for the distribution of those fits was adopted as the main uncertainty of the fit. Also, the standard deviation of the distribution of each variable was considered their main uncertainty.

\begin{figure}[]
   \centering
   \includegraphics[width=\hsize]{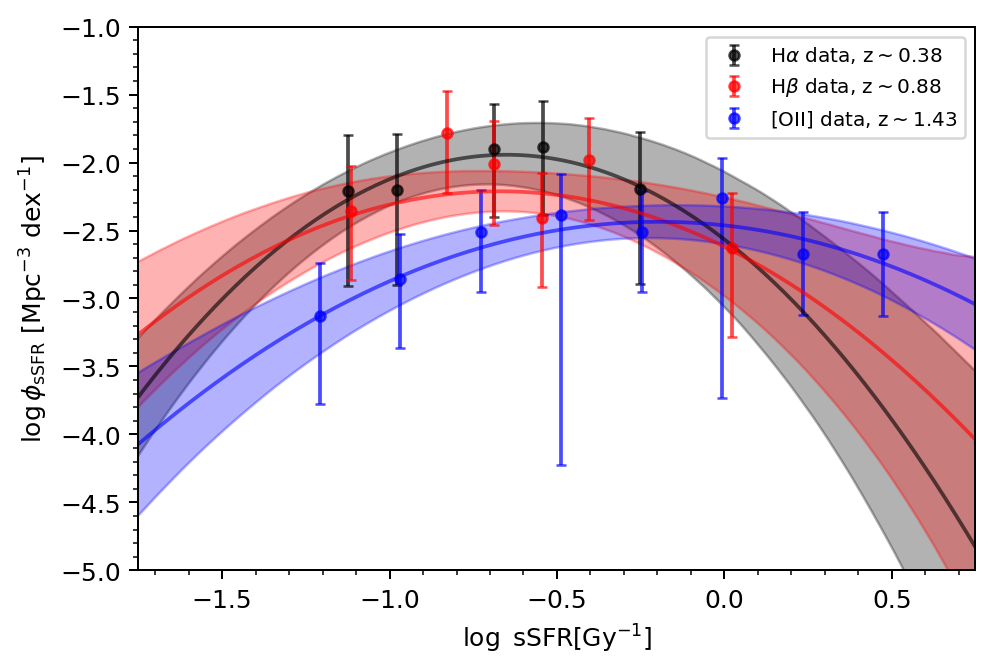}
      \caption{sSFRF of the \ha\, (black dots), \hb\, (red dots), and \oii\, (blue dots) from the OTELO survey. The solid lines represent the log-normal fitting functions for each line (black for \ha, red for \hb, and blue for \oii), following \cite{ilbert2015}'s recipe. The shaded areas represent the propagation of 1$\sigma$ uncertainties of the tabulated fits after $10^5$ Monte Carlo simulations.}
         \label{ssfrt}
 \end{figure}
 
\begin{figure}
    \centering
    \includegraphics[width=\hsize]{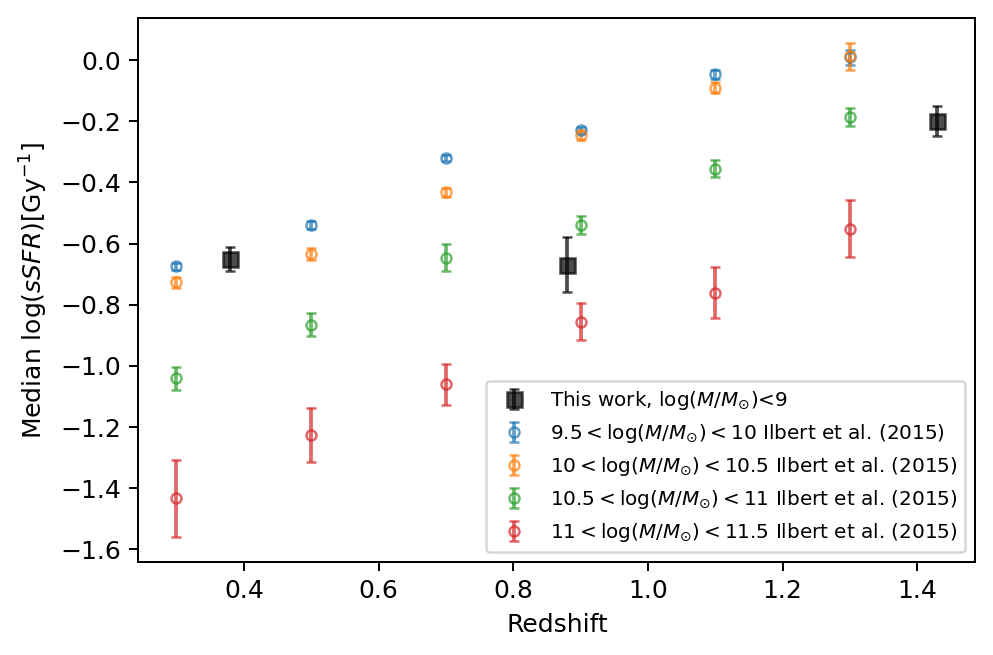}
    \caption{Median value of the sSFR as a function of the redshift for our emitters (filled black squares) and the data from \cite[open circles]{ilbert2015} at different mass bins}
    \label{medianas}
\end{figure}

The sSFRFs for \ha, \hb, and \oii\, emitters and their fits are shown in Fig. \ref{ssfrt}. In Table \ref{funssfr}, we have summarized the parameters of the fit. 
The three functions are somewhat equivalent within the margins of their uncertainties, except for the value of $\log(sSFR^*)$, which is very different for \oii\ emitters compared with \ha\ and \hb. \cite{ilbert2015} demonstrated the evolution of the median of the sSFR function with redshift and with the mass of the emitters. We can also see an evolution of the sSFR with redshift in our data, at least for the \oii\ emitters at $z=1.43$. However, such an evolution is not present from $z=0.38$ to $z=0.90$, as is shown by the sSFRFs of \ha\ and \hb\ emitters. In Fig. \ref{medianas}, we can see the behaviour of the median value of the sSFR as a function of the redshift for our galaxies. We have also represented the values from \cite{ilbert2015} for comparison. The evolution with redshift is more pronounced with \cite{ilbert2015} data. There is also a trend with mass, with higher median values of sSFR for lower masses.  Our results seem to follow the trend with redshift, at least for the \oii\ emitters. However, they present a flat behaviour for lower redshifts. Moreover, considering the masses of the emitters, only the \ha\ ones seem to be located in the same position as the results from \cite{ilbert2015}. Nevertheless, this result should be taken with caution, in particular because the cosmic volumes explored in OTELO and the ones mapped by \cite{ilbert2015} differ by several orders of magnitude.
\noindent

\subsection{The specific star formation density}
\begin{figure}
    \centering
    \includegraphics[width=\hsize]{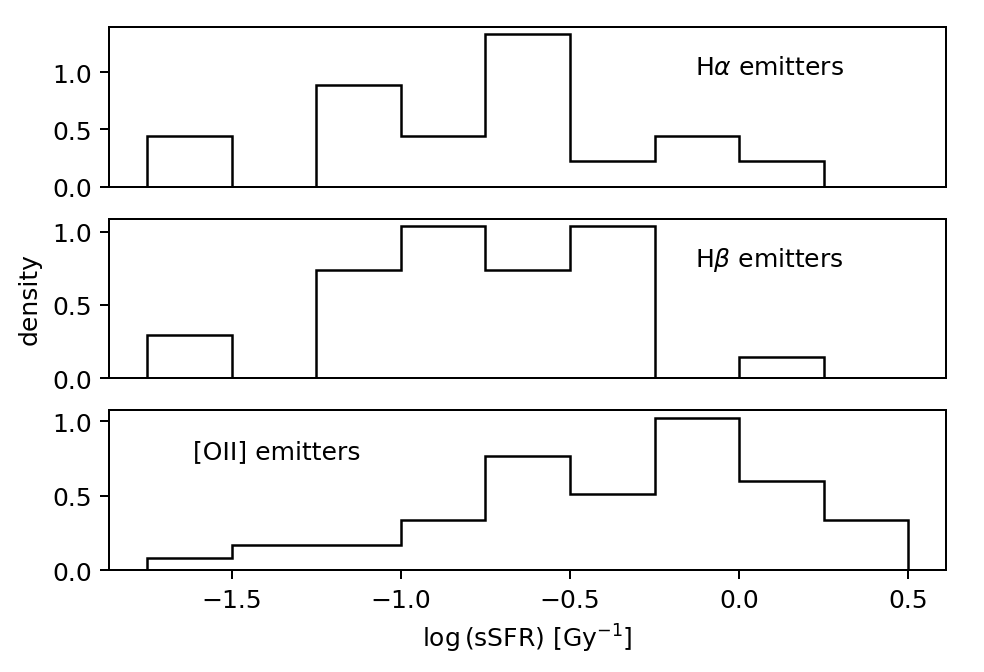}
    \caption{Histogram of the logarithm of the sSFR for the OTELO emitters. Emitters as indicated by the legend.}
    \label{histssfr}
\end{figure}

In a similar fashion as the density of the SFR, calculated in \cite{beliletter}, the density of the sSFR ($\rho_{\mathrm{sSFR}}$) can be calculated by integrating numerically the sSFRF in the following way:

\begin{equation}
    \rho_{\mathrm{sSFR}}=\int \phi(sSFR)\, sSFR\, \mathrm{d}(sSFR)
.\end{equation}
This quantity gives us the sSFR per co-moving volume unit at the selected redshift. In this way, we can take out the possible effect introduced by the different co-moving volumes sampled in our survey.
As the lower integration limit, we selected $\log(sSFR[\mathrm{Gyr^{-1}}])=-2$. This value, suggested by \cite{katsianis2021}, marks the limit of the sSFR below which a galaxy is considered to be a passive object. Figure \ref{histssfr} illustrates the range of the logarithm of the sSFR for the ELGs of the OTELO survey. It can be seen that the sSFR for our sample does not reach that lower limit. This suggests that all our galaxies are actively forming stars, as we would expect from the OTELO survey selection technique based on ELGs. The upper limit of the integral was selected as $\log(sSFR[\mathrm{Gyr^{-1}}])=0.5$, as it is approximately the maximum value of the sSFR for our samples (see Fig. \ref{histssfr}).

\begin{table*}[h]
    \centering
    \caption{Log-normal parameters for the derived sSFRs and the specific star formation density.}
         \begin{tabular}{c c c c c c }
         \hline
         Source & $\langle$z$\rangle$ & $\log \Phi^{*}$ & $\log sSFR^*$ & $\sigma$  & $\log (\rho_{sSFR})$\\
         &  & [${\rm Mpc}^{-3}{\rm dex}^{-1}$] & [Gyr$^{-1}$] & &[Mpc$^{-3}$Gy$^{-1}$]\\
         \hline
        \ha & 0.38 &$-$1.96$\pm$0.04 & $-$0.65$\pm$0.04 & 0.39$\pm$0.10  &  $-$2.26$\pm$0.32\\
        \hb & 0.88 & $-$2.12$\pm$0.08 & $-$0.67$\pm$0.09 & 0.49$\pm$0.09   & $-$2.25$\pm$0.38 \\
        \oii & 1.43 &$-$2.28$\pm$0.04 & $-$0.20$\pm$0.05 & 0.56$\pm$0.05  & $-$1.91$\pm$0.05\\
        \hline
    \end{tabular}
    \label{funssfr}
\end{table*}

\begin{figure}[]
   \centering
   \includegraphics[width=\hsize]{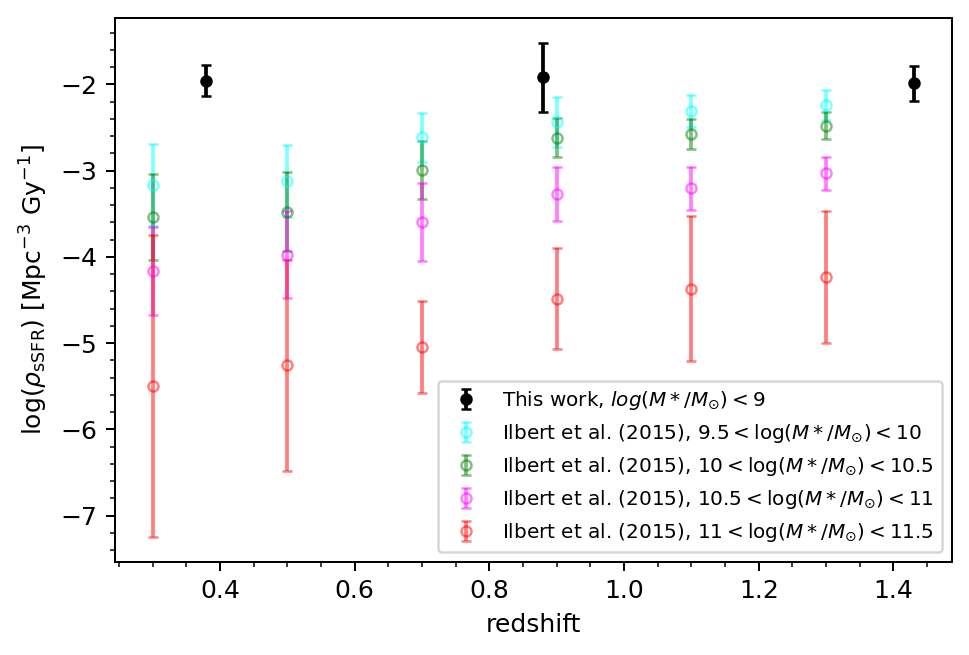}
      \caption{Density of the sSFR as a function of the redshift for our galaxies (filled black dots), and the data from \cite[open circles]{ilbert2015} at different mass bins.}
         \label{ssfrd}
 \end{figure}

\noindent In Fig.\ref{ssfrd}, we have represented the logarithm of the $\rho_{\mathrm{sSFR}}$ as a function of redshift. The open circles represent the values obtained from the integration of the sSFR functions presented in \cite{ilbert2015} for different stellar mass bins. Our results are represented by filled black circles. 
A double trend seems to exist for the $\rho_{\mathrm{sSFR}}$ in the data from \cite{ilbert2015}: the density increases with redshift and decreases with stellar mass. An equivalent double behaviour to this one was previously reported in \cite{ilbert2015}, employing the median value of the sSFR instead of the $\rho_{\mathrm{sSFR}}$. However, our very low-mass galaxies do not show an increase in $\rho_{\mathrm{sSFR}}$ with redshift. In this case, an almost constant value of the density in the $sSFR$ is shown at all redshifts, up to $z\sim1.43$. This happens even for the \oii\ emitters, which present a large difference in the value of $\log(sSFR^*)$.

\section{The stellar mass function and the stellar mass density}\label{sec4}
\subsection{The stellar mass function of the OTELO survey emitters}

For the derived stellar masses, we can follow the same recipe as for the sSFR in Sect. \ref{sec:ssfr}, and define an SMF (also represented by $\phi(M_{\mathrm{s}})$) as the number of galaxies per unit volume and per unit of stellar mass. This function is useful for studying the evolution of the total stellar mass in different volumes of the Universe (\citealt{kikuchihara2020} and references therein).

We also calculate $\phi({M_{\mathrm{s}}})$ in the same way as in Eq. \ref{phi}, simply replacing $\log(sSFR)$ with $\log(M_{\mathrm{s}})$. In this case, we assumed that the detection probability for each source was the same as the one employed in eq. \ref{phi}.

According to \cite{cole2001}, the SMF is well described by a \cite{funcion} function:
\begin{center}
    \begin{equation}
        \phi(M_{\mathrm{s}})\, {\rm d}\,M_{\mathrm{s}} = \phi^\ast (M_{\mathrm{s}}/M^\ast)^\alpha \exp(-M_{\mathrm{s}}/M^\ast)\, {\rm d}\,(M_{\mathrm{s}}/M^\ast),
        \label{sc1}
    \end{equation}
\end{center}
\noindent where $\phi^\ast$ is the density number of galaxies, $M^\ast$ is the characteristic stellar mass, and $\alpha$ is the faint-end slope of the function.
\begin{figure}[]
   \centering
   \includegraphics[width=\hsize]{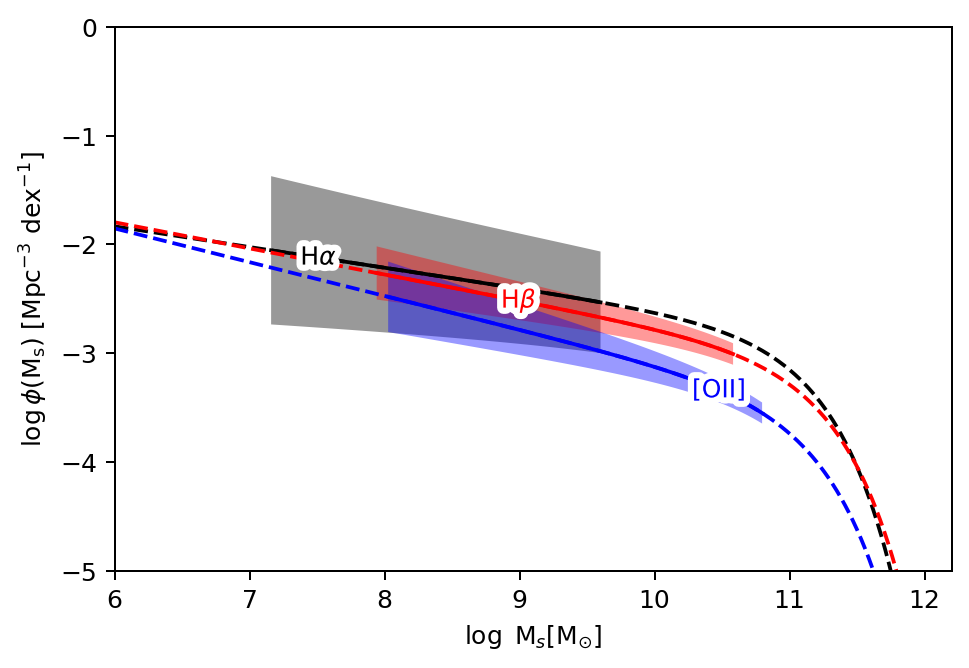}
      \caption{SMF for the \ha, \hb\,, and \oii\ emitters from the OTELO survey, with no mass limit. The solid lines represents the \cite{funcion} fit. Shaded areas were calculated as is indicated in Fig. \ref{ssfrt}.}
         \label{masaha}
 \end{figure}

In Fig. \ref{masaha}, we have represented the SMFs of the emitters from the OTELO survey. The fitting method was the same as the one used for the $\rho_{\mathrm{sSFR}}$ in sec. \ref{sec:ssfr}. It should be taken into account that our galaxies mostly sample the low-mass end of the SMF; thus, the high-mass end is undersampled. So, to minimize the uncertainties, we fixed the value of $M^*$ by interpolating to our redshifts the values of $M^*$ given by \cite{ilbert2013}. The parameters of the SMFs are summarized in Table \ref{funmas}. We obtain a similar $\alpha$ (within uncertainties) to \cite{sobral2014}, pointing to little or no evolution of the faint-end slope. However, considering $\phi^*$, our results for \oii are lower, while for \ha and \hb they are the same within uncertainties.

\begin{table*}[h]
    \centering
    \caption{Schechter parameters for the derived SMFs and the stellar mass function densities integrated for the full sample and the low-mass sample, up to $\log(\rm{M}_s/\rm{M_\sun})=9$ for the SLGs.}
    
    \resizebox{\textwidth}{!}{
         \renewcommand{\arraystretch}{1.5}
         \begin{tabular}{c c c c c c c c c}
         \hline
         Source &  $\langle$z$\rangle$ & $\log \phi^{*}$ & $\log M^*$ (fixed) & $\alpha$ & $\log M_{min}$ & $\log M_{max}$ & $\log (\rho_*)$ (full) & $\log (\rho_*)$ (<=$10^9$M$_\odot$)\\
         &   & [${\rm Mpc}^{-3}{\rm dex}^{-1}$] & [M$_\odot$] & &  [M$_\odot$] &  [M$_\odot$] & [M$_\odot{\rm Mpc}^{-3}$] & [M$_\odot{\rm Mpc}^{-3}$]\\
         \hline
        \ha &  0.38  & $-$3.16$\pm$0.42 & 11.07 & $-$1.19$\pm$0.15 & 7.12 & 9.64 & 7.97$\pm$0.41 & 6.33$\pm$0.60\\
        \hb &  0.88 & $-$3.40$\pm$0.09 & 11.17 & $-$1.24$\pm$0.07 & 7.82 & 10.70 & 7.83$\pm$0.10 & 6.25$\pm$0.20 \\
        \oii & 1.43 & $-$3.79$\pm$0.09 & 11.11 & $-$1.32$\pm$0.10 & 7.90 & 10.93 & 7.44$\pm$0.11 & 6.06$\pm$0.30\\ 
        \hline
    \end{tabular}
    }
   
    \label{funmas}
\end{table*}

\subsection{Stellar mass density}
Equivalent to what was done with the sSFRs, we define the stellar mass density, $\rho_*$, at a certain redshift, as the following integral:
 \begin{equation}
     \rho_*=\int \phi(M_{\mathrm{s}})\, M_{\mathrm{s}}\, \mathrm{d}M_{\mathrm{s}}
. \end{equation}

This integral represents the stellar mass per co-moving volume unit at the explored redshift (\citealt{weaver2023}, \citealt{yu2016}). Usually, the limits are selected as the ones cited by \cite{mimi2016} or \cite{navarro2024}: $10^8\leq M_{*}/\mathrm{M_{\odot}}\leq 10^{13}$. However, taking into account that OTELO survey samples the low-mass regime (66\% of galaxies with $M_{*}/\mathrm{M_{\odot}}< 10^8$, reaching as low as $M_{*}/\mathrm{M_{\odot}}\sim 10^7$), we selected $M_{*}/\mathrm{M_{\odot}}= 10^6$ as the lower limit.

\begin{figure}[]
   \centering
   \includegraphics[width=\hsize]{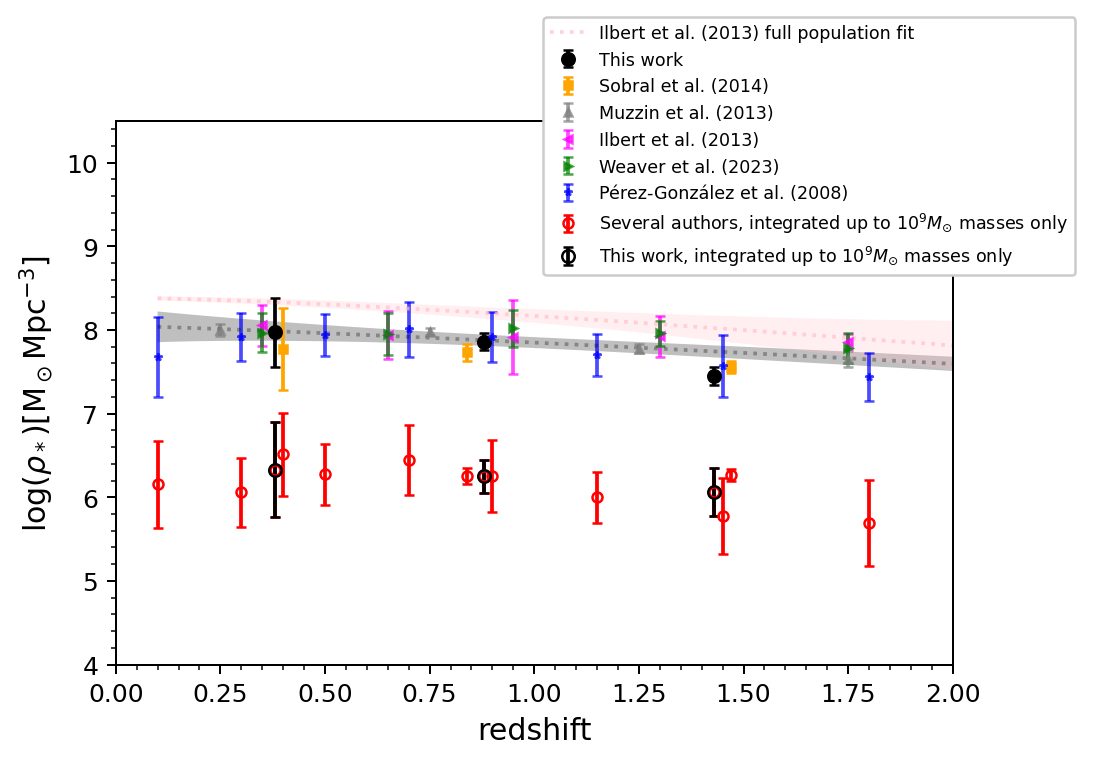}
      \caption{Logarithm of $\rho_*$ in solar masses per cubic megaparsec for the SFGs as a function of redshift up to $z=2$. The filled black circles are our results from the integration of the SMFs for all masses. The open black circles are the same, but integrated only up to $\log(\rm{M}_*[\rm{M}_\odot])=9$, in order to sample only the low-mass galaxy regime. The dotted lines represent the parametric fit by \cite{ilbert2013}. The pink one is the fit for the full population of galaxies (ELGs and quiescent) of \cite{ilbert2013}, and the grey one is the fit for all ELGs from the authors indicated in the legend. Shaded areas were calculated as is indicated in Fig. \ref{ssfrt}.
      The open red circles represent data from \cite{sobral2014} and \cite{perezgonzalez2008}. The remaining symbols and colour coding are reported in the legend.}
         \label{denmass}
 \end{figure}

If we take into account the SFGs and integrate the SMFs presented in table \ref{funmas}, we obtain the results of Fig. \ref{denmass}. We have also included data from literature (\citealt{sobral2014}, \citealt{muzzin2013}, \citealt{ilbert2013}, \citealt{weaver2023}, and \citealt{perezgonzalez2008}). We have also calculated $\rho_*$ for low-mass galaxies only up to $10^9\mathrm{M}_{\odot}$ (represented by open circles). To generate the red symbols of Fig. \ref{denmass}, we selected data from authors who sampled the low-mass end of the SMF (\citealt{perezgonzalez2008} and \citealt{sobral2014}).

Following \cite{ilbert2013}, we fitted the evolution of the stellar mass density to this parametric function:
\begin{equation}
    \rho_*(z)=a\times e^{-b z^c}
    \label{ajilbert}
.\end{equation}
In Table \ref{ajusteilbert}, we have summarized the parameters of Eq. \ref{ajilbert} for our data combined with the results reported from \cite{sobral2014}, \cite{muzzin2013}, \cite{ilbert2013}, \cite{weaver2023}, and \cite{perezgonzalez2008} for ELGs only (called ‘compilation’ in the table), as well as the results from \cite{ilbert2013} for a population of ELGs and quiescent galaxies. When comparing both fits, we can see that the main difference resides in the parameter $\log(a)$, the value of the function at $z=0$. The exponent, $c$, and the constant, $b$, are compatible with each other.  
\begin{table*}[]
    \centering
    \caption{Parameters derived for the fitting of  eq. \ref{ajilbert}.}
    \begin{tabular}{c|c|c|c}
    \hline
    Source & $\log(a)$ & b & c\\
    \hline
    \cite{ilbert2013} full population & 8.39$\pm$0.02 & 0.50$\pm$0.19 & 1.41$\pm$0.40\\
    OTELO ELGs + compilation & 8.05$\pm$0.08 & 0.46$\pm$0.18 & 1.18$\pm$0.24\\
    \hline
    \end{tabular}  
    \label{ajusteilbert}
\end{table*}

On the other hand, the fit for the low-mass galaxies with Eq. \ref{ajilbert} did not converge. Nevertheless, our data follows the trend presented by other authors integrating from $10^{6}\mathrm{M_{\odot}}$ only up to $10^{9}\mathrm{M_{\odot}}$. It seems that for low-mass galaxies the trend is somewhat compatible with a constant value, but we have to take into account that we are only sampling up to $z\sim2$. As happened with $\rho_{\mathrm{sSFR}}$, the difference in $\log(\phi^*)$ for our \oii\ emitters does not seem to have too much influence on $\rho_*$, at least for the limited redshift range that we are studying and for integrating up to $10^{9}\mathrm{M_{\odot}}$.

To study this apparent difference in behaviour, we followed the method presented in \cite{fontanot2009}. We created bins in stellar mass and integrated the SMFs within those bins. This will generate a $\rho_*$ for each bin. From Table \ref{resgal}, we can see that the lowest mass we are able to achieve is about $\log (M_{min}/\mathrm{M_{\odot}})\sim7.12$; we are able to create bins starting at $\log(M_{s}/\mathrm{M_{\odot}})=7$.

So, in the end we created five bins in $\log(M_{s}/\mathrm{M_{\odot}})$, from (7,8] to (11,12]. To check the possible evolution of the $\rho_*$, \cite{fontanot2009} calculated a linear regression of $\log(\rho_*)$ with the redshift, for each mass bin, and studied the variation in the slope. For bins (7,8] and (8,9], we employed only data from this study, and for the rest of the bins, we also included the data from \cite{sobral2014} and \cite{perezgonzalez2008}. Those works were selected because they were the ones that reached lower values for the stellar mass.

\begin{figure}
    \centering
    \includegraphics[width=\hsize]{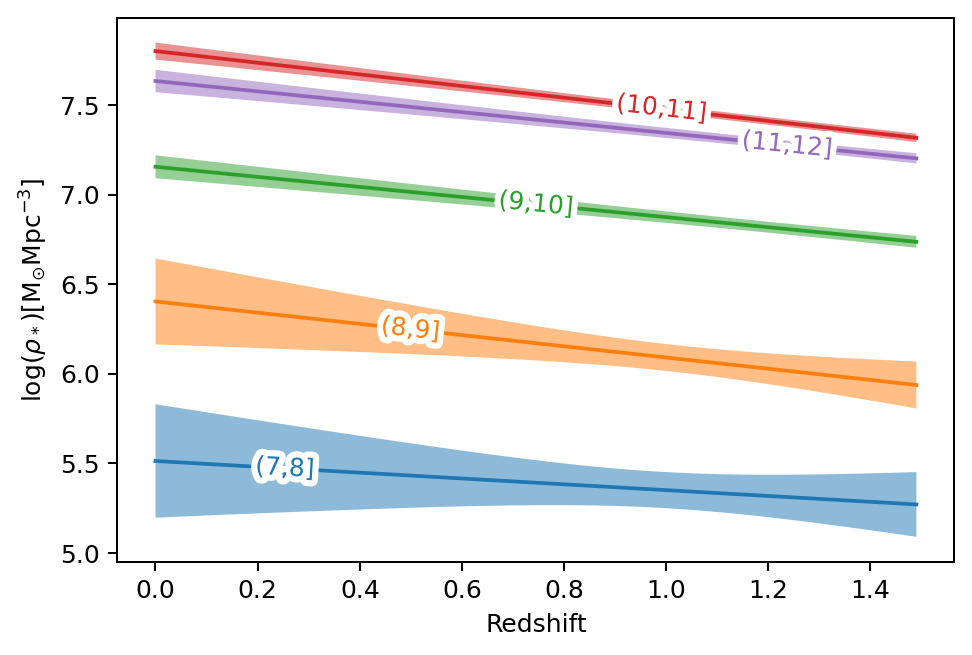}
    \caption{Fits of the stellar mass function densities as a function of redshift in bins of galaxy stellar mass. Bins (6,7], (7,8], and (8,9] were generated with data from this work exclusively. Bin (9,10] also includes data from \cite{sobral2014}. In bins (10,11] and (11,12], there is also added data from \cite{perezgonzalez2008}. The shaded area represents the propagation of 1$\sigma$ uncertainties.}
    \label{ajusteall}
\end{figure}

\begin{figure}
    \centering
    \includegraphics[width=\hsize]{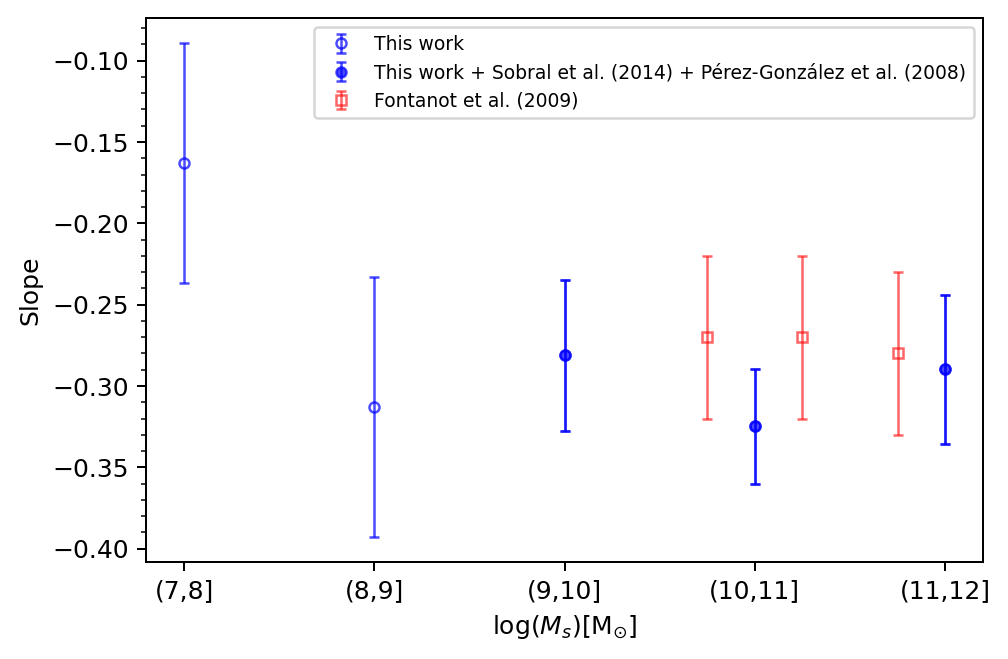}
    \caption{Slopes derived for the fits presented in Fig. \ref{ajusteall}.}
    \label{pendientes}
\end{figure}

In Fig. \ref{ajusteall}, we have represented the fitted regressions to $\log(\rho_*)$ versus $z$. The shaded areas represent the propagation of 1$\sigma$ uncertainties and were calculated as in Fig. \ref{ssfrt}. In Fig. \ref{pendientes}, we present the slopes of the regressions as a function of the bins. A different slope behaviour for the bin with the lowest mass seems to exist, with a higher value than the rest of the bins. From the (8,9] bin, the results are consistent with the ones from \cite{fontanot2009}, in which it reaches a minimum of about -0.30 for the higher masses.

\cite{fontanot2009} proposed that the downsizing implies that more massive galaxies should be assembled at higher redshifts and in shorter times than low-mass galaxies. This is then translated to $\rho_*$ with a different growth rate for each bin versus the redshift. However, \cite{fontanot2009} found that the slopes for all the bins with masses between $10^9<M_s/\mathrm{M_{\odot}}<10^{12}$ were similar, with no clear differences among them. This result is replicated in Fig.  \ref{pendientes}, in which our data points are compatible with those from \cite{fontanot2009}. We can assume that the slope obtained for the bin that is lowest in mass may be an indication of a faster evolution of the more massive galaxies compared with very low-mass ones. However, it should be noted that this is only valid for the lowest masses, and the uncertainties present are large enough that more observations are required in order to confirm this.

\section{The star formation history}\label{sec5}
To study the SFHs of the ELGs of our sample, we performed a SED fitting to the galaxies employing the Code Investigating GALaxy Emission (CIGALE), (\citealt{boquien2019}). We employed the bands in the range from 1530\AA\ (GALEX far-ultraviolet filter) to 45000\AA\ (IRAC 4.5$\mu$m filter). We selected a delayed SFH, a \cite{bruzual2003} synthesis model with stellar metallicities from 0.004 to 0.05, and \cite{calzetti2000}'s dust attenuation model with a colour excess, $E(B-V)$, from 0 to 1.1.

The delayed SFH selected is a simple way to generate the SFH. It is based on a delayed exponential SFR and a later exponential burst to simulate the most recent process in star formation. It can be written, following the parameterization proposed by \cite{malek2018}, as:
\begin{equation}
SFR\propto
    \left\{\begin{matrix}
        SFR_{\mathrm{delayed}} & t<t_{0}  \\
        SFR_{\mathrm{delayed}} + SFR_{\mathrm{burst}}& t> t_{0}  \\
    \end{matrix}\right.
    \label{mainsfr}
,\end{equation}
where:
\begin{equation}
    SFR_{\mathrm{delayed}}\propto \frac{t}{\tau_{\mathrm{main}}^2} \times \exp{(-t/\tau_\mathrm{{main}})}
    \label{eq.delayed}
\end{equation}
and
\begin{equation}
    SFR_{\mathrm{burst}}\propto \exp-(t-t_0)/\tau_{\mathrm{burst}}
    \label{eq.burst}
.\end{equation}
Here, $\tau_{\mathrm{main}}$ and $\tau_{\mathrm{burst}}$ are the e-folding times for the main starburst population and the late starburst, respectively.

After the fitting, the galaxies with large uncertainties in the determination of the stellar mass, as well as galaxies with a low number of bands ($<4$) in the SED, were rejected.
\subsection{Stellar mass comparison}
\begin{figure}
    \centering
    \includegraphics[width=\hsize]{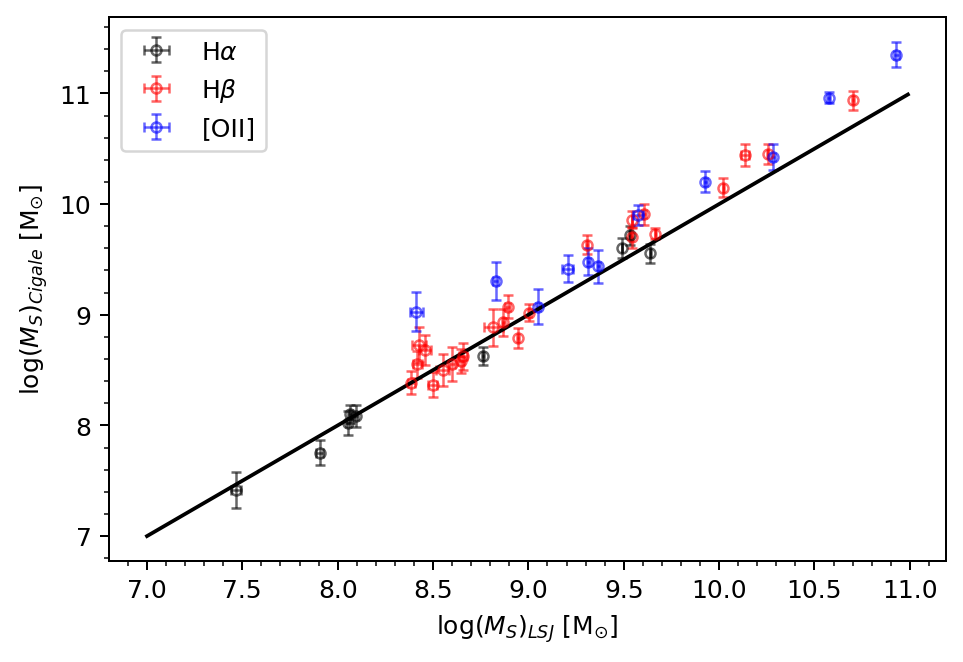}
    \caption{Comparison between the stellar masses determined by the \cite{lopezsanjuan2019} method of \cite{jakub2020} and the stellar masses obtained by the CIGALE fit in this work. The black line indicates a 1:1 equivalence. Emitters are indicated by the legend.}
    \label{CIGALEmass}
\end{figure}

As a first check of our new SED fits, we can compare the stellar masses obtained from \cite{jakub2020} with \cite{lopezsanjuan2019}'s recipe with those given by CIGALE. In Fig. \ref{CIGALEmass}, we have represented both mass determinations. The black line represents the 1:1 relationship. We can see that both quantities are equivalent within the uncertainty limit (0.14\,dex at 1$\sigma$), although CIGALE determination is slightly larger at higher masses, with a larger dispersion (0.17\,dex at 1$\sigma$).
Taking into account this relationship, we are able to consider any determination of the stellar mass interchangeably. However, we use the mass determination of \cite{jakub2020}. 

\subsection{The $\tau_{\mathrm{main}}$ factor}
The value of $\tau_{\mathrm{main}}$ (e-fold time) from Eq. \ref{eq.delayed} indicates how fast the SFR decays with time. A large enough value of $\tau_{\mathrm{main}}$ may simulate a continuous SFR, while a lower value generates a burst that decays faster with time.

\begin{figure}
    \centering
    \includegraphics[width=\hsize]{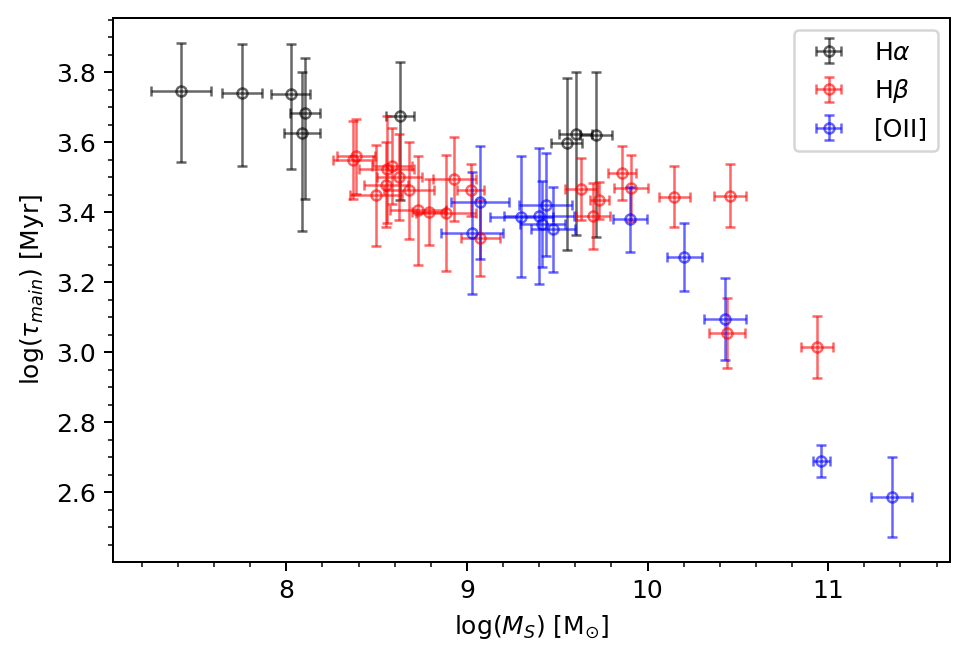}
    \caption{Value of the logarithm of the parameter $\tau_{\mathrm{main}}$ from Eq. \ref{eq.delayed} as a function of stellar mass for the OTELO emitters. Symbols are as in the legend.}
    \label{tau}
\end{figure}

In Fig. \ref{tau}, we have represented the e-fold time for the main stellar burst as a function of the stellar mass for galaxies in our sample. It can be observed that there is a slight trend with the stellar mass: higher-mass galaxies, over $\log(M_{\mathrm{s}}/M_{\odot})>10$, have lower values of $\tau_{\mathrm{main}}$ compared with lower-mass galaxies. Moreover, galaxies with $\log(M_{\mathrm{s}}/M_{\odot})<10$ exhibit a constant behaviour within the uncertainties. 

This seems to indicate that for low-mass galaxies the star-forming process tends to be more constant compared with the most massive galaxies in our sample, whose SFRs decay faster with time. Although we may need a larger number of galaxies with masses over $10^{10}$M$_{\odot}$ in order to extract a clear conclusion, these results seem to be consistent with the previous ones from Secs. \ref{sec3} and \ref{sec4}.

On the other hand, the \ha\ emitters seem to have a slightly different behaviour compared with \hb\ and \oii\ emitters. However, the uncertainties present and the low number of \ha\ ELGs prevent us from drawing definitive conclusions.
\subsection{The $D4000$ parameter}
One of the derived parameters from the CIGALE fit is the $D4000$ ratio. \cite{bruzual1983} defined it as the ratio between the fluxes in the range of 4050\,$\mathrm{\AA}$--4250\,$\mathrm{\AA}$ and 3750\,$\mathrm{\AA}$--3950\,$\mathrm{\AA}$. 
\noindent
This index depends on the metallicity, but it also correlates with the ratio of the present to past SFR. In \cite{brinchmann2004}, it was shown that the sSFR depends on $D4000$, with lower values of sSFR having higher values of $D4000$.
\begin{figure}
    \centering
    \includegraphics[width=\hsize]{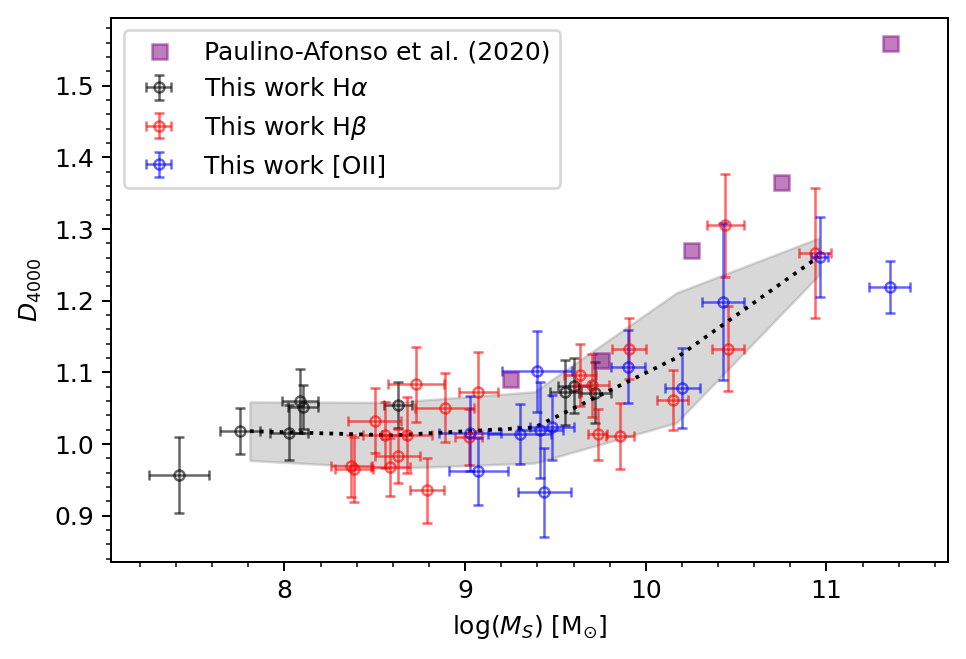}
    \caption{Parameter $D\,4000$ as a function of stellar mass. The dotted black line represents the median value for bins of 0.75\,dex. The grey area represents the propagation of 1$\sigma$ uncertainties. Purple squares represent the data from \cite{paulino2020}. The rest of symbols are as in the legend.}
    \label{d4000}
\end{figure}
In Fig. \ref{d4000}, we show the relationship of $D4000$ with the stellar mass of the galaxies of our sample. We calculated the median value of our data by employing bins of 0.75\,dex (dotted black line). We have also included data from \cite{paulino2020} at redshift $z\sim0.8$. However, in that paper, they employ $D_n4000$ (which uses a somewhat different integration limit than $D4000$). Nevertheless, they found that both $D4000$ and $D_n4000$ correlate very well, so it makes no difference whether we use one ratio or the other.

An almost constant value can be seen in Fig. \ref{d4000} for the median of $D4000$ for stellar masses of $\log(M_{\mathrm{s}}/\mathrm{M_{\odot}})<10$, and an increase in $D4000$ can be seen at higher masses.  
Moreover, the stellar mass where the transition occurs is more or less the same as the one shown in Fig. \ref{tau}. On the other hand, our results seem to agree with those from \cite{paulino2020} for the range $9<\log(M_{\mathrm{s}}/\mathrm{M_{\odot}})<10$. There is a difference at higher masses that can be attributed to our lack of emitters in such a regime.


\section{Summary and conclusions}\label{sec6}

    In this work, we present the results of a study of the ELGs of the OTELO survey, covering the redshifts 0.38 (\ha), 0.88 (\hb), and 1.43 (\oii). Due to the characteristics of the survey, the selected emitters lie mainly in the low-mass regime, with limits between $7.0<\log(M_*/\mathrm{M_{\odot}})<10.5$.

    We calculated the sSFRF for OTELO emitters and fitted them to a log-normal function, following \cite{ilbert2015}. We represented the median sSFR as a function of redshift and found no evolution from \ha\ and \hb\ emitters, up to a redshift of $z=0.88$. We compared these results with \cite{ilbert2015} and found differences in their behaviour with the redshift and masses. These differences can be explained by the large differences in the co-moving volumes sampled by both surveys and the CV implied. We also calculated the $\rho_{\mathrm{sSFR}}$ for OTELO emitters, integrating the sSFRF. We compared them with the results from \cite{ilbert2015}. Our results show a flat behaviour of the $\rho_{\mathrm{sSFR}}$ as a function of redshift for galaxies with $\log(M_*/\mathrm{M_{\odot}})<9$. On the other hand, the results from \cite{ilbert2015} show a trend for higher-mass galaxies, with higher values of $\rho_{\mathrm{sSFR}}$ with increasing redshift.

    We calculated the SMF for the OTELO emitters. Then, each SMF was fitted to a \cite{funcion} function. After that, we calculated $\rho_*$ by integrating the SMFs employing different integration limits.

    We cannot discard an evolution of the sSFRF or the SMF with redshift in our data, mainly due to differences in the fit parameters for \oii\ emitters, compared with \ha\ and \hb. However, this evolution does not have a large influence on the determination of the densities $\rho_*$ and $\rho_{\mathrm{sSFR}}$.

    Following \cite{ilbert2013}, $\rho_*$ was fitted to a parametric function. The results obtained agree with those from the literature. Compared with the results from \cite{ilbert2013}, in which a full population of galaxies (emitters and quiescent) was considered, the values of almost all the parameters of the fit are the same within the uncertainties. The only difference lies in the ordinate at origin, with a large value for the full population of \cite{ilbert2013}, compared with a pure ELG sample from several authors (this work plus \citealt{sobral2014} and \citealt{perezgonzalez2008}).

    Following the method presented by \cite{fontanot2009}, we integrated $\rho_*$ in bins of $\log(M_*)$ and represented them as a function of the redshift. We then fitted the results to a linear regression. We selected bins between (7,8] to (11,12]. We assumed that the low-mass-end slope of the SMFs is well defined in our sample, so we extended our results to stellar mass regimes of (7,8].

    We found a change in the slope of the fit with the mass bin. For the (7,8] bin, there is a higher slope. The slope then decreases to more negative values with higher bins in mass. According to \cite{fontanot2009}, this change in slope is an indication of downsizing, and it is only perceived at very low masses.

    We performed a fit of the SEDs of the emitters employing the CIGALE code in order to study the behaviour of the SFH. We found that the value of the e-folding time of the main starburst population ($\tau_{\mathrm{main}}$) is dependent on the stellar mass: a constant larger value for low-mass galaxies, and a decreasing lower value for high-mass galaxies. This seems to imply a more constant star formation process with time for low-mass galaxies and a faster decline in SFR for higher-mass galaxies.

    We represented the $D4000$ ratio as a function of the stellar mass. We found, on average, a constant value for low-mass galaxies and an increase for high-mass galaxies. This may indicate a change in the sSFR, which is higher in value and which has a constant behaviour with redshift for low-mass galaxies, as was presented in Sect. \ref{sec3}.

Our analysis suggests that the main drivers of the star formation process change at a critical stellar mass of $\log(M_{\mathrm{s}}/\mathrm{M_{\odot}})>9.5$. This critical mass has been reported as well in the well-known mass-metallicity relationship (\citealt{tremonti2004}), according to which the gas metallicity of galaxies tends to flatten for higher stellar masses.

Taken together, all these results seem to indicate that low-mass galaxies tend to maintain a more or less constant star formation process for longer periods of time, while more massive galaxies experience episodes of vigorous star formation that are shorter in time duration. However, we must take into account that this is only valid up to a redshift of $z\sim1.5$.

\begin{acknowledgements}

The authors wish to thank the anonymous referee for her/his feed-back and useful suggestions. 

B.C. and J.C. ~acknowledge the suppor of the Spanish Ministry of Science, Innovation and Universities through the project PID--2021--122544NB--C41. 

A.B. and M.S.P. ~acknowledge the support of the Spanish Ministry of Science, Innovation and Universities through the project PID--2021--122544NB--C43. 

J.N. ~acknowledge the support of 
the National Science Centre, Poland through the SONATA BIS grant 2018/30/E/ST9/00208, 2021/41/N/ST9/02662
and 2023/49/B/ST9/00066 and the support of the Polish National Agency for Academic Exchange (NAWA) Bekker  grant BPN/BEK/2023/1/00271. 

J.G. ~acknowledge the support of he Spanish Ministry of Science, Innovation and Universities through the project PID--2021--123417OB--I00. 

J.I.G-S. ~acknowledge the support of he Spanish Ministry of Science, Innovation and Universities through the project PID--2021--122544NB--C44. 

M.A.L.L. ~acknowledges support from the Spanish grant PID2021--123417OB--I00, and the Ramón y Cajal program funded by the Spanish Government (RYC2020--029354--I). 

M.C. ~acknowledges support from the Spanish grants PID-2021--122544NB--C41 and PID2022--136598NB--C33 funded by MCIN/AEI/10.13039/501100011033 and by “ERDF A way of making Europe”. 

E.J.A. ~has been partially funded by the Spanish MCIN/AEI grant PID2022--136640NB--C21 and acknowledges financial support from the Severo Ochoa grant CEX2021--001131--S funded by MCIN/AEI/ 10.13039/501100011033. 

J.A.D. ~acknowledges the support of UNAM--PAPIIT grant IN116325. 

M.P. ~acknowledges the support of the Spanish Ministry of Science, Innovation and Universities
through the project PID2022--140871NB--C21, the Severo Ochoa grant CEX2021--515001131--S funded by MCIN/AEI/10.13039/501100011033, and support from the Space Science and Geospatial Institute (SSGI) funded through the Ministry of Innovation and Technology (MInT). 

B.C. wishes to thank Carlota Leal \'Alvarez by her support during the development of this paper.  

This article is based on observations made with the Gran Telescopio Canarias (GTC) at Roque de los Muchachos Observatory of the Instituto de Astrof\'isica de Canarias on the island of La Palma. 

This study makes use of data from AEGIS, a multi-wavelength sky survey conducted with the Chandra, GALEX, Hubble, Keck, CFHT, MMT, Subaru, Palomar, Spitzer, VLA, and other telescopes, and is supported in part by the NSF, NASA, and the STFC. 

Based  on  observations  obtained  with  MegaPrime/MegaCam,  a  joint  project  of the  CFHT  and CEA/IRFU, at the Canada--France--Hawaii Telescope (CFHT) which is operated by the National Research Council (NRC) of Canada, the Institut National des Science de l'Univers of the Centre National de la Recherche Scientifique (CNRS) of France, and the University of Hawaii.  This work is based in part on data products produced at Terapix available at the Canadian Astronomy Data Centre as part of the Canada-France-Hawaii Telescope Legacy Survey, a collaborative project of NRC and CNRS. 

Based on observations obtained with WIRCam, a joint project of CFHT, Taiwan, Korea, Canada, France, at the Canada--France--Hawaii Telescope (CFHT), which is operated by the National Research Council (NRC) of Canada, the Institute National des Sciences de l'Univers of the Centre National de la Recherche Scientifique of France, and the University of Hawaii. This work is based in part on data products produced at TERAPIX, the WIRDS (WIRcam Deep Survey) consortium, and the Canadian Astronomy Data Centre. This research was supported by a grant from the Agence Nationale de la Recherche ANR-07-BLAN-0228. 
\end{acknowledgements}

%

\bibliographystyle{aa} 
\bibliography{biblio} 

\begin{thebibliography}{51}
\expandafter\ifx\csname natexlab\endcsname\relax\def\natexlab#1{#1}\fi

\bibitem[{{Arnouts} {et~al.}(1999){Arnouts}, {Cristiani}, {Moscardini},
  {Matarrese}, {Lucchin}, {Fontana}, \& {Giallongo}}]{arnouts1999}
{Arnouts}, S., {Cristiani}, S., {Moscardini}, L., {et~al.} 1999, \mnras, 310,
  540

\bibitem[{{Bertin} \& {Arnouts}(1996)}]{bertin1996}
{Bertin}, E. \& {Arnouts}, S. 1996, \aaps, 117, 393

\bibitem[{{Bongiovanni} {et~al.}(2019){Bongiovanni}, {Ram{\'o}n-P{\'e}rez},
  {P{\'e}rez Garc{\'\i}a}, {Cepa}, {Cervi{\~n}o}, {Nadolny}, {P{\'e}rez
  Mart{\'\i}nez}, {Alfaro}, {Casta{\~n}eda}, {de Diego}, {Ederoclite},
  {Fern{\'a}ndez-Lorenzo}, {Gallego}, {Gonz{\'a}lez}, {Gonz{\'a}lez-Serrano},
  {Lara-L{\'o}pez}, {Oteo G{\'o}mez}, {Padilla Torres}, {Pintos-Castro},
  {Povi{\'c}}, {S{\'a}nchez-Portal}, {Jones}, {Bland-Hawthorn}, \&
  {Cabrera-Lavers}}]{otelo}
{Bongiovanni}, {\'A}., {Ram{\'o}n-P{\'e}rez}, M., {P{\'e}rez Garc{\'\i}a},
  A.~M., {et~al.} 2019, \aap, 631, A9

\bibitem[{{Bongiovanni} {et~al.}(2020){Bongiovanni}, {Ram{\'o}n-P{\'e}rez},
  {P{\'e}rez Garc{\'\i}a}, {Cervi{\~n}o}, {Cepa}, {Nadolny}, {P{\'e}rez
  Mart{\'\i}nez}, {Alfaro}, {Casta{\~n}eda}, {Cedr{\'e}s}, {de Diego},
  {Ederoclite}, {Fern{\'a}ndez-Lorenzo}, {Gallego}, {de Jes{\'u}s
  Gonz{\'a}lez}, {Gonz{\'a}lez-Serrano}, {Lara-L{\'o}pez}, {Oteo G{\'o}mez},
  {Padilla Torres}, {Pintos-Castro}, {Povi{\'c}}, {S{\'a}nchez-Portal}, {Heath
  Jones}, {Bland-Hawthorn}, \& {Cabrera-Lavers}}]{bongio2020}
{Bongiovanni}, {\'A}., {Ram{\'o}n-P{\'e}rez}, M., {P{\'e}rez Garc{\'\i}a},
  A.~M., {et~al.} 2020, \aap, 635, A35

\bibitem[{{Boquien} {et~al.}(2019){Boquien}, {Burgarella}, {Roehlly}, {Buat},
  {Ciesla}, {Corre}, {Inoue}, \& {Salas}}]{boquien2019}
{Boquien}, M., {Burgarella}, D., {Roehlly}, Y., {et~al.} 2019, \aap, 622, A103

\bibitem[{{Brinchmann} {et~al.}(2004){Brinchmann}, {Charlot}, {White},
  {Tremonti}, {Kauffmann}, {Heckman}, \& {Brinkmann}}]{brinchmann2004}
{Brinchmann}, J., {Charlot}, S., {White}, S.~D.~M., {et~al.} 2004, \mnras, 351,
  1151

\bibitem[{{Bruzual} \& {Charlot}(2003)}]{bruzual2003}
{Bruzual}, G. \& {Charlot}, S. 2003, \mnras, 344, 1000

\bibitem[{{Bruzual A.}(1983)}]{bruzual1983}
{Bruzual A.}, G. 1983, \apj, 273, 105

\bibitem[{{Calzetti} {et~al.}(2000){Calzetti}, {Armus}, {Bohlin}, {Kinney},
  {Koornneef}, \& {Storchi-Bergmann}}]{calzetti2000}
{Calzetti}, D., {Armus}, L., {Bohlin}, R.~C., {et~al.} 2000, \apj, 533, 682

\bibitem[{{Cedr{\'e}s} {et~al.}(2021{\natexlab{a}}){Cedr{\'e}s}, {Bongiovanni},
  {Cervi{\~n}o}, {Nadolny}, {Cepa}, {de Diego}, {P{\'e}rez Garc{\'\i}a},
  {Gallego}, {Lara-L{\'o}pez}, {S{\'a}nchez-Portal}, {Gonz{\'a}lez-Serrano},
  {Alfaro}, {Navarro Mart{\'\i}nez}, {P{\'e}rez Mart{\'\i}nez}, {Gonz{\'a}lez},
  {Padilla Torres}, {Casta{\~n}eda}, \& {Gonz{\'a}lez}}]{beli2021}
{Cedr{\'e}s}, B., {Bongiovanni}, {\'A}., {Cervi{\~n}o}, M., {et~al.}
  2021{\natexlab{a}}, \aap, 649, A73

\bibitem[{{Cedr{\'e}s} {et~al.}(2021{\natexlab{b}}){Cedr{\'e}s},
  {P{\'e}rez-Garc{\'\i}a}, {P{\'e}rez-Mart{\'\i}nez}, {Cervi{\~n}o}, {Gallego},
  {Bongiovanni}, {Cepa}, {Navarro Mart{\'\i}nez}, {Nadolny}, {Lara-L{\'o}pez},
  {S{\'a}nchez-Portal}, {Alfaro}, {de Diego}, {Gonz{\'a}lez-Otero}, {Jes{\'u}s
  Gonz{\'a}lez}, {Ignacio Gonz{\'a}lez-Serrano}, \& {Padilla
  Torres}}]{beliletter}
{Cedr{\'e}s}, B., {P{\'e}rez-Garc{\'\i}a}, A.~M., {P{\'e}rez-Mart{\'\i}nez},
  R., {et~al.} 2021{\natexlab{b}}, \apjl, 915, L17

\bibitem[{{Cole} {et~al.}(2001){Cole}, {Norberg}, {Baugh}, {Frenk},
  {Bland-Hawthorn}, {Bridges}, {Cannon}, {Colless}, {Collins}, {Couch},
  {Cross}, {Dalton}, {De Propris}, {Driver}, {Efstathiou}, {Ellis},
  {Glazebrook}, {Jackson}, {Lahav}, {Lewis}, {Lumsden}, {Maddox}, {Madgwick},
  {Peacock}, {Peterson}, {Sutherland}, \& {Taylor}}]{cole2001}
{Cole}, S., {Norberg}, P., {Baugh}, C.~M., {et~al.} 2001, \mnras, 326, 255

\bibitem[{{Cowie} {et~al.}(1996){Cowie}, {Songaila}, {Hu}, \&
  {Cohen}}]{cowie1996}
{Cowie}, L.~L., {Songaila}, A., {Hu}, E.~M., \& {Cohen}, J.~G. 1996, \aj, 112,
  839

\bibitem[{{Desprez} {et~al.}(2023){Desprez}, {Picouet}, {Moutard}, {Arnouts},
  {Sawicki}, {Coupon}, {Gwyn}, {Chen}, {Huang}, {Golob}, {Furusawa}, {Ikeda},
  {Paltani}, {Cheng}, {Hartley}, {Hsieh}, {Ilbert}, {Kauffmann}, {McCracken},
  {Shuntov}, {Tanaka}, {Toft}, {Tresse}, \& {Weaver}}]{desprez2023}
{Desprez}, G., {Picouet}, V., {Moutard}, T., {et~al.} 2023, \aap, 670, A82

\bibitem[{{Fontanot} {et~al.}(2009){Fontanot}, {De Lucia}, {Monaco},
  {Somerville}, \& {Santini}}]{fontanot2009}
{Fontanot}, F., {De Lucia}, G., {Monaco}, P., {Somerville}, R.~S., \&
  {Santini}, P. 2009, \mnras, 397, 1776

\bibitem[{{Guzm{\'a}n} {et~al.}(1997){Guzm{\'a}n}, {Gallego}, {Koo},
  {Phillips}, {Lowenthal}, {Faber}, {Illingworth}, \& {Vogt}}]{guzman1997}
{Guzm{\'a}n}, R., {Gallego}, J., {Koo}, D.~C., {et~al.} 1997, \apj, 489, 559

\bibitem[{{Henry} {et~al.}(2013){Henry}, {Scarlata}, {Dom{\'\i}nguez},
  {Malkan}, {Martin}, {Siana}, {Atek}, {Bedregal}, {Colbert}, {Rafelski},
  {Ross}, {Teplitz}, {Bunker}, {Dressler}, {Hathi}, {Masters}, {McCarthy}, \&
  {Straughn}}]{henry2013}
{Henry}, A., {Scarlata}, C., {Dom{\'\i}nguez}, A., {et~al.} 2013, \apjl, 776,
  L27

\bibitem[{{Hopkins}(2004)}]{Hopkins2004}
{Hopkins}, A.~M. 2004, \apj, 615, 209

\bibitem[{{Ilbert} {et~al.}(2015){Ilbert}, {Arnouts}, {Le Floc'h}, {Aussel},
  {Bethermin}, {Capak}, {Hsieh}, {Kajisawa}, {Karim}, {Le F{\`e}vre}, {Lee},
  {Lilly}, {McCracken}, {Michel-Dansac}, {Moutard}, {Renzini}, {Salvato},
  {Sanders}, {Scoville}, {Sheth}, {Silverman}, {Smol{\v{c}}i{\'c}},
  {Taniguchi}, \& {Tresse}}]{ilbert2015}
{Ilbert}, O., {Arnouts}, S., {Le Floc'h}, E., {et~al.} 2015, \aap, 579, A2

\bibitem[{{Ilbert} {et~al.}(2006){Ilbert}, {Arnouts}, {McCracken},
  {Bolzonella}, {Bertin}, {Le F{\`e}vre}, {Mellier}, {Zamorani}, {Pell{\`o}},
  {Iovino}, {Tresse}, {Le Brun}, {Bottini}, {Garilli}, {Maccagni}, {Picat},
  {Scaramella}, {Scodeggio}, {Vettolani}, {Zanichelli}, {Adami}, {Bardelli},
  {Cappi}, {Charlot}, {Ciliegi}, {Contini}, {Cucciati}, {Foucaud}, {Franzetti},
  {Gavignaud}, {Guzzo}, {Marano}, {Marinoni}, {Mazure}, {Meneux}, {Merighi},
  {Paltani}, {Pollo}, {Pozzetti}, {Radovich}, {Zucca}, {Bondi}, {Bongiorno},
  {Busarello}, {de La Torre}, {Gregorini}, {Lamareille}, {Mathez}, {Merluzzi},
  {Ripepi}, {Rizzo}, \& {Vergani}}]{ilbert2006}
{Ilbert}, O., {Arnouts}, S., {McCracken}, H.~J., {et~al.} 2006, \aap, 457, 841

\bibitem[{{Ilbert} {et~al.}(2013){Ilbert}, {McCracken}, {Le F{\`e}vre},
  {Capak}, {Dunlop}, {Karim}, {Renzini}, {Caputi}, {Boissier}, {Arnouts},
  {Aussel}, {Comparat}, {Guo}, {Hudelot}, {Kartaltepe}, {Kneib}, {Krogager},
  {Le Floc'h}, {Lilly}, {Mellier}, {Milvang-Jensen}, {Moutard}, {Onodera},
  {Richard}, {Salvato}, {Sanders}, {Scoville}, {Silverman}, {Taniguchi},
  {Tasca}, {Thomas}, {Toft}, {Tresse}, {Vergani}, {Wolk}, \&
  {Zirm}}]{ilbert2013}
{Ilbert}, O., {McCracken}, H.~J., {Le F{\`e}vre}, O., {et~al.} 2013, \aap, 556,
  A55

\bibitem[{{Katsianis} {et~al.}(2021){Katsianis}, {Xu}, {Yang}, {Luo}, {Cui},
  {Dav{\'e}}, {Lagos}, {Zheng}, \& {Zhao}}]{katsianis2021}
{Katsianis}, A., {Xu}, H., {Yang}, X., {et~al.} 2021, \mnras, 500, 2036

\bibitem[{{Kennicutt}(1998)}]{kennicutt1998}
{Kennicutt}, Jr., R.~C. 1998, \araa, 36, 189

\bibitem[{{Kewley} {et~al.}(2004){Kewley}, {Geller}, \& {Jansen}}]{kewley2004}
{Kewley}, L.~J., {Geller}, M.~J., \& {Jansen}, R.~A. 2004, \aj, 127, 2002

\bibitem[{{Kikuchihara} {et~al.}(2020){Kikuchihara}, {Ouchi}, {Ono},
  {Mawatari}, {Chevallard}, {Harikane}, {Kojima}, {Oguri}, {Bruzual}, \&
  {Charlot}}]{kikuchihara2020}
{Kikuchihara}, S., {Ouchi}, M., {Ono}, Y., {et~al.} 2020, \apj, 893, 60

\bibitem[{{Kinney} {et~al.}(1996){Kinney}, {Calzetti}, {Bohlin}, {McQuade},
  {Storchi-Bergmann}, \& {Schmitt}}]{kinney1996}
{Kinney}, A.~L., {Calzetti}, D., {Bohlin}, R.~C., {et~al.} 1996, \apj, 467, 38

\bibitem[{{Kroupa}(2001)}]{kroupaimf}
{Kroupa}, P. 2001, \mnras, 322, 231

\bibitem[{{Lara-L{\'o}pez} {et~al.}(2010){Lara-L{\'o}pez}, {Cepa},
  {Bongiovanni}, {P{\'e}rez Garc{\'\i}a}, {Ederoclite}, {Casta{\~n}eda},
  {Fern{\'a}ndez Lorenzo}, {Povi{\'c}}, \& {S{\'a}nchez-Portal}}]{maritza2010}
{Lara-L{\'o}pez}, M.~A., {Cepa}, J., {Bongiovanni}, A., {et~al.} 2010, \aap,
  521, L53

\bibitem[{{Lehnert} {et~al.}(2015){Lehnert}, {van Driel}, {Le Tiran}, {Di
  Matteo}, \& {Haywood}}]{lehnert}
{Lehnert}, M.~D., {van Driel}, W., {Le Tiran}, L., {Di Matteo}, P., \&
  {Haywood}, M. 2015, \aap, 577, A112

\bibitem[{{L{\'o}pez-Sanjuan} {et~al.}(2019){L{\'o}pez-Sanjuan},
  {D{\'\i}az-Garc{\'\i}a}, {Cenarro}, {Fern{\'a}ndez-Soto}, {Viironen},
  {Molino}, {Ben{\'\i}tez}, {Crist{\'o}bal-Hornillos}, {Moles}, {Varela},
  {Arnalte-Mur}, {Ascaso}, {Castander}, {Cervi{\~n}o}, {Gonz{\'a}lez Delgado},
  {Husillos}, {M{\'a}rquez}, {Masegosa}, {Del Olmo}, {Povi{\'c}}, \&
  {Perea}}]{lopezsanjuan2019}
{L{\'o}pez-Sanjuan}, C., {D{\'\i}az-Garc{\'\i}a}, L.~A., {Cenarro}, A.~J.,
  {et~al.} 2019, \aap, 622, A51

\bibitem[{{Madau} \& {Dickinson}(2014)}]{madau2014}
{Madau}, P. \& {Dickinson}, M. 2014, \araa, 52, 415

\bibitem[{{Ma{\l}ek} {et~al.}(2018){Ma{\l}ek}, {Buat}, {Roehlly}, {Burgarella},
  {Hurley}, {Shirley}, {Duncan}, {Efstathiou}, {Papadopoulos}, {Vaccari},
  {Farrah}, {Marchetti}, \& {Oliver}}]{malek2018}
{Ma{\l}ek}, K., {Buat}, V., {Roehlly}, Y., {et~al.} 2018, \aap, 620, A50

\bibitem[{{Marchesini} {et~al.}(2009){Marchesini}, {van Dokkum}, {F{\"o}rster
  Schreiber}, {Franx}, {Labb{\'e}}, \& {Wuyts}}]{marchesini2009}
{Marchesini}, D., {van Dokkum}, P.~G., {F{\"o}rster Schreiber}, N.~M., {et~al.}
  2009, \apj, 701, 1765

\bibitem[{{Muzzin} {et~al.}(2013){Muzzin}, {Marchesini}, {Stefanon}, {Franx},
  {McCracken}, {Milvang-Jensen}, {Dunlop}, {Fynbo}, {Brammer}, {Labb{\'e}}, \&
  {van Dokkum}}]{muzzin2013}
{Muzzin}, A., {Marchesini}, D., {Stefanon}, M., {et~al.} 2013, \apj, 777, 18

\bibitem[{{Nadolny} {et~al.}(2020){Nadolny}, {Lara-L{\'o}pez}, {Cervi{\~n}o},
  {Bongiovanni}, {Cepa}, {de Diego}, {P{\'e}rez Garc{\'\i}a}, {P{\'e}rez
  Mart{\'\i}nez}, {S{\'a}nchez-Portal}, {Alfaro}, {Casta{\~n}eda}, {Gallego},
  {Gonz{\'a}lez}, {Gonz{\'a}lez-Serrano}, {Padilla Torres}, {Pintos-Castro}, \&
  {Povi{\'c}}}]{jakub2020}
{Nadolny}, J., {Lara-L{\'o}pez}, M.~A., {Cervi{\~n}o}, M., {et~al.} 2020, \aap,
  636, A84

\bibitem[{{Navarro-Carrera} {et~al.}(2024){Navarro-Carrera}, {Rinaldi},
  {Caputi}, {Iani}, {Kokorev}, \& {van Mierlo}}]{navarro2024}
{Navarro-Carrera}, R., {Rinaldi}, P., {Caputi}, K.~I., {et~al.} 2024, \apj,
  961, 207

\bibitem[{{Navarro Mart{\'\i}nez} {et~al.}(2021){Navarro Mart{\'\i}nez},
  {P{\'e}rez-Garc{\'\i}a}, {P{\'e}rez-Mart{\'\i}nez}, {Cervi{\~n}o}, {Gallego},
  {Bongiovanni}, {Barrufet}, {Nadolny}, {Cedr{\'e}s}, {Cepa}, {Alfaro},
  {Casta{\~n}eda}, {de Diego}, {Gonz{\'a}lez-Otero}, {Jes{\'u}s Gonz{\'a}lez},
  {Gonz{\'a}lez-Serrano}, {Lara-L{\'o}pez}, {Padilla Torres}, \&
  {S{\'a}nchez-Portal}}]{rocio2021}
{Navarro Mart{\'\i}nez}, R., {P{\'e}rez-Garc{\'\i}a}, A.~M.,
  {P{\'e}rez-Mart{\'\i}nez}, R., {et~al.} 2021, \aap, 653, A24

\bibitem[{{Paulino-Afonso} {et~al.}(2020){Paulino-Afonso}, {Sobral}, {Darvish},
  {Ribeiro}, {Smail}, {Best}, {Stroe}, \& {Cairns}}]{paulino2020}
{Paulino-Afonso}, A., {Sobral}, D., {Darvish}, B., {et~al.} 2020, \aap, 633,
  A70

\bibitem[{{P{\'e}rez-Gonz{\'a}lez} {et~al.}(2008){P{\'e}rez-Gonz{\'a}lez},
  {Rieke}, {Villar}, {Barro}, {Blaylock}, {Egami}, {Gallego}, {Gil de Paz},
  {Pascual}, {Zamorano}, \& {Donley}}]{perezgonzalez2008}
{P{\'e}rez-Gonz{\'a}lez}, P.~G., {Rieke}, G.~H., {Villar}, V., {et~al.} 2008,
  \apj, 675, 234

\bibitem[{{Picouet} {et~al.}(2023){Picouet}, {Arnouts}, {Le Floc'h}, {Moutard},
  {Kraljic}, {Ilbert}, {Sawicki}, {Desprez}, {Laigle}, {Schiminovich}, {de la
  Torre}, {Gwyn}, {McCracken}, {Dubois}, {Dav{\'e}}, {Toft}, {Weaver},
  {Shuntov}, \& {Kauffmann}}]{picouet2023}
{Picouet}, V., {Arnouts}, S., {Le Floc'h}, E., {et~al.} 2023, \aap, 675, A164

\bibitem[{{Ram{\'o}n-P{\'e}rez} {et~al.}(2019){Ram{\'o}n-P{\'e}rez},
  {Bongiovanni}, {P{\'e}rez Garc{\'\i}a}, {Cepa}, {Lara-L{\'o}pez}, {de Diego},
  {Alfaro}, {Casta{\~n}eda}, {Cervi{\~n}o}, {Fern{\'a}ndez-Lorenzo}, {Gallego},
  {Gonz{\'a}lez}, {Gonz{\'a}lez-Serrano}, {Nadolny}, {Oteo G{\'o}mez},
  {P{\'e}rez Mart{\'\i}nez}, {Pintos-Castro}, {Povi{\'c}}, \&
  {S{\'a}nchez-Portal}}]{marina2019}
{Ram{\'o}n-P{\'e}rez}, M., {Bongiovanni}, {\'A}., {P{\'e}rez Garc{\'\i}a},
  A.~M., {et~al.} 2019, \aap, 631, A10

\bibitem[{{Sales} {et~al.}(2022){Sales}, {Wetzel}, \& {Fattahi}}]{sales2022}
{Sales}, L.~V., {Wetzel}, A., \& {Fattahi}, A. 2022, Nature Astronomy, 6, 897

\bibitem[{{Schechter}(1976)}]{funcion}
{Schechter}, P. 1976, \apj, 203, 297

\bibitem[{{Simmonds} {et~al.}(2024){Simmonds}, {Tacchella}, {Hainline},
  {Johnson}, {McClymont}, {Robertson}, {Saxena}, {Sun}, {Witten}, {Baker},
  {Bhatawdekar}, {Boyett}, {Bunker}, {Charlot}, {Curtis-Lake}, {Egami},
  {Eisenstein}, {Hausen}, {Maiolino}, {Maseda}, {Scholtz}, {Williams},
  {Willott}, \& {Witstok}}]{simmonds2024}
{Simmonds}, C., {Tacchella}, S., {Hainline}, K., {et~al.} 2024, \mnras, 527,
  6139

\bibitem[{{Sobral} {et~al.}(2014){Sobral}, {Best}, {Smail}, {Mobasher},
  {Stott}, \& {Nisbet}}]{sobral2014}
{Sobral}, D., {Best}, P.~N., {Smail}, I., {et~al.} 2014, \mnras, 437, 3516

\bibitem[{{Somerville} {et~al.}(2004){Somerville}, {Lee}, {Ferguson},
  {Gardner}, {Moustakas}, \& {Giavalisco}}]{somerville2004}
{Somerville}, R.~S., {Lee}, K., {Ferguson}, H.~C., {et~al.} 2004, \apjl, 600,
  L171

\bibitem[{{Song} {et~al.}(2016){Song}, {Finkelstein}, {Ashby}, {Grazian}, {Lu},
  {Papovich}, {Salmon}, {Somerville}, {Dickinson}, {Duncan}, {Faber}, {Fazio},
  {Ferguson}, {Fontana}, {Guo}, {Hathi}, {Lee}, {Merlin}, \&
  {Willner}}]{mimi2016}
{Song}, M., {Finkelstein}, S.~L., {Ashby}, M. L.~N., {et~al.} 2016, \apj, 825,
  5

\bibitem[{{Tremonti} {et~al.}(2004){Tremonti}, {Heckman}, {Kauffmann},
  {Brinchmann}, {Charlot}, {White}, {Seibert}, {Peng}, {Schlegel}, {Uomoto},
  {Fukugita}, \& {Brinkmann}}]{tremonti2004}
{Tremonti}, C.~A., {Heckman}, T.~M., {Kauffmann}, G., {et~al.} 2004, \apj, 613,
  898

\bibitem[{{Weaver} {et~al.}(2023){Weaver}, {Davidzon}, {Toft}, {Ilbert},
  {McCracken}, {Gould}, {Jespersen}, {Steinhardt}, {Lagos}, {Capak}, {Casey},
  {Chartab}, {Faisst}, {Hayward}, {Kartaltepe}, {Kauffmann}, {Koekemoer},
  {Kokorev}, {Laigle}, {Liu}, {Long}, {Magdis}, {McPartland}, {Milvang-Jensen},
  {Mobasher}, {Moneti}, {Peng}, {Sanders}, {Shuntov}, {Sneppen}, {Valentino},
  {Zalesky}, \& {Zamorani}}]{weaver2023}
{Weaver}, J.~R., {Davidzon}, I., {Toft}, S., {et~al.} 2023, \aap, 677, A184

\bibitem[{{Wilkins} {et~al.}(2008){Wilkins}, {Trentham}, \&
  {Hopkins}}]{wilkins2008}
{Wilkins}, S.~M., {Trentham}, N., \& {Hopkins}, A.~M. 2008, \mnras, 385, 687

\bibitem[{{Yu} \& {Wang}(2016)}]{yu2016}
{Yu}, H. \& {Wang}, F.~Y. 2016, \apj, 820, 114

\end{thebibliography}

\end{document}